\documentclass[12pt]{article}
\usepackage{graphicx}
\usepackage{amsmath}
\usepackage{bbm}
\usepackage{amsfonts}
\usepackage{enumerate}
\usepackage[lofdepth,lotdepth]{subfig}
\usepackage{natbib}
\usepackage{threeparttable}
\usepackage[geometry]{ifsym}
\usepackage[nolists]{endfloat}
\usepackage{url}

\newcommand{\bs}{\boldsymbol}

\newcommand{\iid}{\stackrel{\mathrm{i.i.d.}}{\sim}}

\newtheorem{thm}{Theorem}
\newtheorem{lem}{Lemma}

\newtheorem{prop}{Proposition}
\newtheorem{cor}{Corollary}

\def\urltilda{\kern -.15em\lower .7ex\hbox{\~{}}\kern .04em}
\def\urldot{\kern -.10em.\kern -.10em}
\def\urlhttp{http\kern -.10em\lower -.1ex\hbox{:}\kern -.12em\lower 0ex\hbox{/}\kern -.18em\lower 0ex\hbox{/}}

\begin{document}

\title{\bfseries Stochastic Volatility Regression for \\Functional Data Dynamics
\author{Bin Zhu and David B. Dunson$^*$}
}

\date{}

\maketitle
\let\thefootnote\relax\footnotetext{$^*$Bin Zhu is Tenure-Track Principal Investigator, Biostatistics Branch,
Division of Cancer Epidemiology and Genetics,
National Cancer Institute,
National Institutes of Health, Rockville, MD 20852 (Email: \emph{bin.zhu@nih.gov}). David B. Dunson is Professor, Department of Statistical Science, Duke University, Durham, NC 27708, (Email: \emph{dunson@stat.duke.edu}). }

\centerline{\textbf{Abstract}}
\noindent
Although there are many methods for functional data analysis (FDA), little emphasis is put on characterizing variability among volatilities of individual functions.  In particular, certain individuals exhibit erratic swings in their trajectory while other individuals have more stable trajectories.  There is evidence of such volatility heterogeneity in blood pressure trajectories during pregnancy, for example, and reason to suspect that volatility is a biologically important feature.  Most FDA models implicitly assume similar or identical smoothness of the individual functions, and hence can lead to misleading inferences on volatility and an inadequate representation of the functions.  We propose a novel class of FDA models characterized using hierarchical stochastic differential equations.  We model the derivatives of a mean function and deviation functions using Gaussian processes, while also allowing covariate dependence including on the volatilities of the deviation functions.  Following a Bayesian approach to inference, a Markov chain Monte Carlo algorithm is used for posterior computation.  The methods are tested on simulated data and applied to blood pressure trajectories during pregnancy. 

\newpage
\section{Introduction}
\label{sec:intro}
Multi-subject functional data arise frequently in many fields of research, including epidemiology, clinical trials and environmental health.  Sequential observations are collected over time for multiple subjects, and can be treated as being generated from a smooth trajectory contaminated with noise.  There are a rich variety of methods available for characterizing variability and covariate dependence in functional data ranging from hierarchical basis expansions to functional principal components analysis (FPCA).   In defining models for functional data and in interpreting variability in trajectories, either unexplained or due to covariates, the emphasis has been on differences in the level and local trends.  Dynamic features, such as velocity, acceleration and especially volatility, are also important but receive much less attention.  

Analysis of functional data dynamics studies temporal changes in trajectories and effects of covariates on these changes. For example, \citet{wang2008modeling} used linear differential equations to model price velocity and acceleration in eBay auctions and explored the auction subpopulation effect. \citet{muller2010empirical} modeled the velocity of 
online auction bids using empirical stochastic differential equations with time-varying coefficients and a smooth drift process. \citet{zhu2011semiparametric} inferred the rate functions of prostate-specific antigen profiles using the proposed semiparametric stochastic velocity model, in which rate functions are regarded as realizations of Ornstein-Uhlenbeck processes dependent on covariates of interest. 

This article investigates a different dynamic feature, the volatility, which measures the conditional variance of trajectory changes over an infinitesimal time interval. We propose a stochastic volatility regression (SVR) model with Gaussian process (GP) priors used for the group mean and subject specific deviation functions through stochastic differential equations (SDEs). We further accommodate inference on covariate effects on volatility through allowing the diffusion term of SDEs for deviation functions to depend on covariates.
Although volatility has been extensively studied through stochastic volatility (SV) models in finance \citep{heston1993closed,jacquier2002bayesian,shephard2005stochastic, barndorff2011financial}, the setting, model specifications and data features are distinct from ours.  SV models typically focus on a single volatility process which is time-varying and associated with a price process for high-frequency finance data.  More relevant is the literature on multivariate SV models; for recent references, refer to \citet{loddo2011selection}, \citet{van2011estimation}, \citet{ishihara2012efficient} and \citet{durante2012locally}.

This setting differs from ours in that the focus is on multivariate time series modeling instead of functional data analysis, with interest in the joint volatility dynamics over time for the different assets.  In contrast, we are interested in studying variation across individuals in a time-constant subject-specific volatility; that is, certain subjects may have very smooth trajectories while other subjects have erratic trajectories.  It is our conjecture that such volatility heterogeneity is common in biomedical settings, but is overlooked in analyzing data with models that implicitly prescribe a single level of smoothness for all subjects.  As data are sparse and irregularly spaced in most studies, it is not surprising such behavior is overlooked.  However, the volatility in a biomarker may be as important or more important than the overall level and trend in the biomarker.  We provide motivation through the following longitudinal blood pressure data set.

\begin{figure}
\centering
\subfloat[]{\label{fig:1a}\includegraphics[width=0.48\textwidth,angle=270]{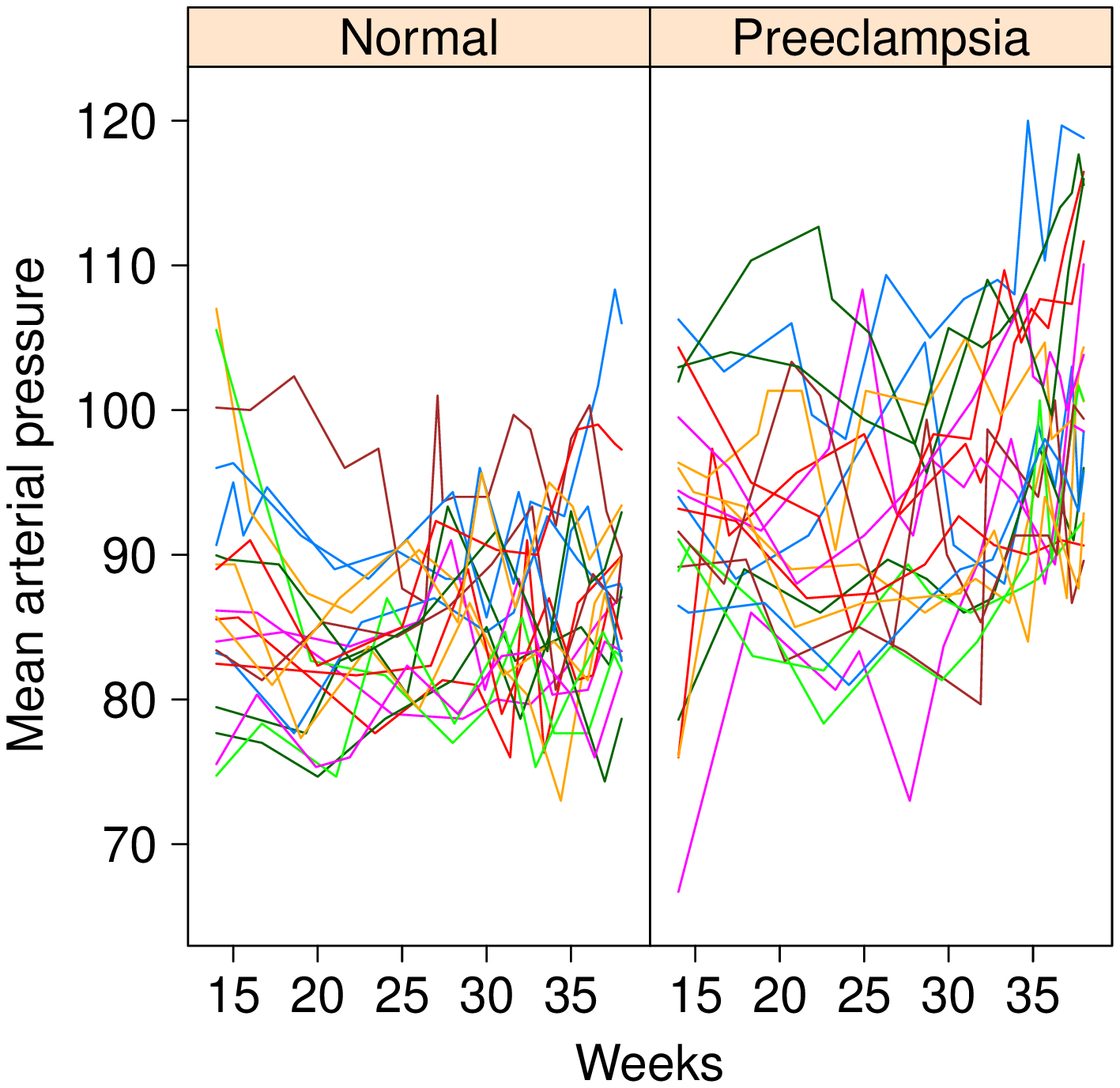}}
\subfloat[]{\label{fig:1b}\includegraphics[width=0.46\textwidth,angle=270]{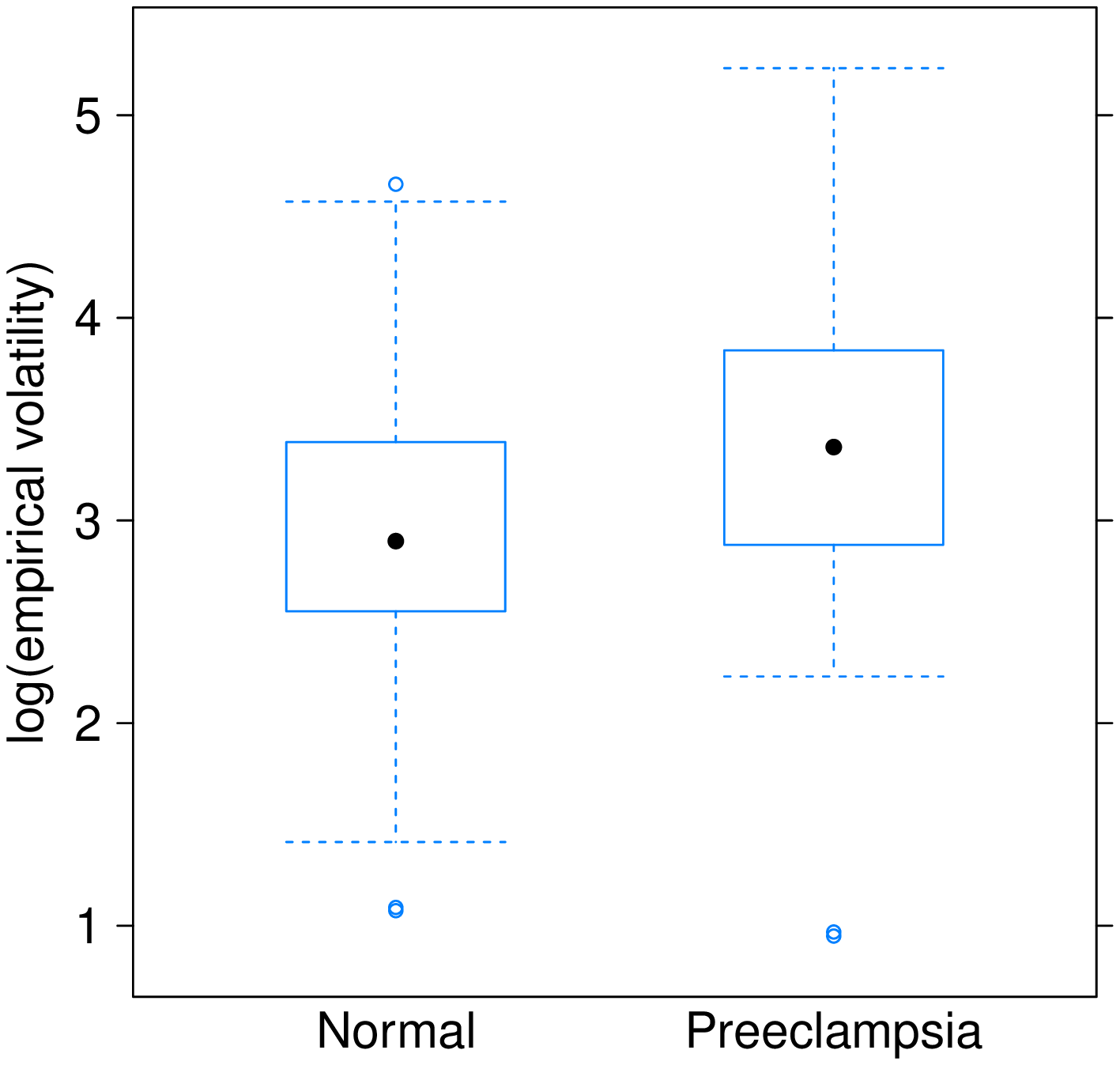}}
\caption{(a) Mean arterial pressure (MAP) trajectories for twenty randomly selected normal women and women with preeclampsia; (b) Log-transformed empirical volatilities for women in the normal group and preeclampsia group. $Y_{ij}$ denotes blood pressure for the $i$th woman at time $t_{ij}$,  and  
 $U_{ij} = Y_{ij} - \bar{Y}_{j}$ is the deviation from the group mean blood pressure $\bar{Y}_{j}$. The empirical volatility measures the fluctuation of trajectories empirically and is defined as $\frac{1}{n_i}\sum_{j=1}^{n_i-1}\frac{( U_{i,j+1}-U_{i,j} )^2}{t_{i,j+1}-t_{i,j}}$ with $n_i$ the number of observations for the $i$th woman.     \label{fig:empiricalVol}}  
\end{figure}

The Healthy Pregnancy, Healthy Baby Study \citep[HPHB, ][]{miranda2009environmental} collected longitudinal blood pressure (BP) measurements for pregnant women. Blood pressures are measured at irregularly spaced times during the second and third trimesters with the number of measurements per subject varying from 9 to 19. We are interested in estimating subject-specific volatilities of BP trajectories and in identifying covariates associated with the volatility. Figure \ref{fig:empiricalVol}\subref{fig:1a} plots mean arterial pressure (MAP) trajectories for twenty randomly selected normal women and women with preeclampsia, respectively. Clearly the MAP trajectories among the preeclampsia group are more wiggly than the ones in the normal group, which is also implied by  boxplots of log-transformed empirical volatilities in Figure \ref{fig:empiricalVol}\subref{fig:1b}. To explore volatility differences among various groups in addition to preeclampsia, we apply normal linear regression for log-transformed empirical volatilities with the covariates race, mother's age, obesity, preeclampsia, parity and smoking. The results suggest that preeclampsia and smoking (p-values 0.0005 and 0.002) are associated with empirical volatility. This is a two-stage approach, which is useful as an exploratory tool but ignores measurement errors and uncertainty in volatility estimation.    

Additionally, empirical volatilities in Figure \ref{fig:empiricalVol}\subref{fig:1b} are heterogeneous even within the normal or preeclampsia group. This heterogeneity will be largely omitted when we apply FDA methods with identical or similar smoothness for individual functions within a group. Consequently, the wiggly trajectories will be over-smoothed while the smooth trajectories will be under-smoothed. We can potentially estimate the individual trajectories separately but it is well known that borrowing of information will dramatically improve performance for sparse functional data.  In addition, separate estimation does not allow for inferences on covariate effects and unexplained variability in volatility.  

As for the clinical question addressed, the previous FDA methods mainly focus on the shift of blood pressure level and ignore examining the volatility of blood pressure, which measures the haemodynamic stability and is crucial for cardiovascular health. For example, a recent study shows that blood pressure stability rather than blood pressure level is associated with increased survival among patients on hemodialysis \citep{raimann2012blood}.
For the HPHB study, we observe that preeclampsia is commonly accompanied by blood pressure over-swinging. The joint effect of high blood pressure level and large volatility may lead to the adverse birth outcomes, such as low birth weight and preterm birth. 

The remainder of the article is organized as follows. Section \ref{sec:SVR}
specifies the SVR model and discusses its properties. Section  \ref{sec:posterior} develops an efficient Markov chain Monte Carlo algorithm  for posterior inference. Section \ref{sec:simu} presents simulation studies and the proposed method is applied to a real dataset in Section \ref{sec:app}. Finally, section \ref{sec:dis} contains concluding remarks and future possible extensions.

\section{Stochastic Volatility Regression Model}
\label{sec:SVR}

\subsection{The Model Specification}
Suppose that $Y_i(t)$, $i=1,2,\dots, m$, is the observation of the $i$th subject at  time $t \in \mathcal{T}_i = \{t_{i,1}, t_{i,2}, \cdots, t_{i,n_i} < t_U\}$ with $\mathcal{T}_i$ the set of observation times before time $t_U$ for the $i$th subject. We specify an observation equation for $Y_i(t)$ as
\begin{equation}
\label{eq:obs}
Y_i(t) = M_{k_i}(t) + U_{i}(t) + \varepsilon_i(t),
\end{equation} 
where $Y_i(t)$ is contaminated by the measurement error $\varepsilon_i(t) $ following a one-dimensional normal distribution with mean $0$ and variance $\sigma^2_\varepsilon$. Assuming the $i$th subject belongs to the ${k_i}$th group (e.g. by race or treatment) with ${k_i} \in \{1, 2, \cdots, g\}$,   we include a ${k_i}$th group mean function $M_{k_i}(t) = \textsf{E}\{Y_i(t) \mid M_{k_i}(t)\}$ in the observation equation. In addition, the trajectory of the $i$th subject will unlikely coincide with $M_{k_i}(t)$ and therefore the departure from $M_{k_i}(t)$ is addressed and represented by the subject-specific deviation function $U_{i}(t)$  with $\textsf{E}\{U_i(t) \}=0$.

The volatility of the $i$th subject is defined as the conditional variance of the $(q-1)$th order derivative of $U_i(t)$ over an infinitesimal time interval. Namely, we denote the volatility  $\sigma^2_{U_i} =\mathop {\lim }\limits_{h \to 0} h^{-1} \textsf{E}\left[\left\{D^{q-1}U_i(t+h) - D^{q-1}U_i(t)\right\}^2\mid D^{q-1}U_i(t) \right]$ with differential operator $D^q=\frac{d^q}{dt^q}$. As volatility approaches zero, $U_i(t)$ would be a roughly flat line. In contrast, increasing the value of volatility would lead to a more wiggly $U_i(t)$  with a larger magnitude of fluctuation around $M_{k_i}(t)$.  

We specify Gaussian process priors for $M_{k_i}(t)$ and $U_i(t)$ using 
SDEs which incorporate the group and individual volatilities $\sigma^2_{M_{k_i}}$ and $\sigma^2_{U_i}$   : 
\begin{align}
\label{eq:sde_M_k}
D^pM_{k_i}(t) &=\sigma_{M_{k_i}} \dot{W}_{k_i}(t), \\
\label{eq:sde_U_i}
D^{q} U_i(t) &= \sigma_{U_i}\dot{W}^\prime_i(t),
\end{align}
%$L^p_{k_i} = \frac{d}{dt} \frac{1}{N_{k_i, 1}(t)}
%             \frac{d}{dt} \frac{1}{N_{k_i, 2}(t)}
%             \cdots
%             \frac{d}{dt} \frac{1}{N_{k_i, p}(t)}
%$
%and
%$L^q = \frac{d}{dt} \frac{1}{V_{1}(t)}
%       \frac{d}{dt} \frac{1}{V_{2}(t)}
%       \cdots
%       \frac{d}{dt} \frac{1}{V_{q}(t)}
%$
%are the differential operators with $N_{k_i, 1}(t)$, $N_{k_i, 2}(t)$, $\cdots$, $N_{k_i, p}(t)$ and $V_{1}(t)$, $V_{2}(t)$, $\cdots$, $V_{q}(t)$ the nontrivial continuous functions incorporating our prior knowledge regarding the shape of the functions;  
where $p, q \in \mathbb{N}\geq 1$ and  $\sigma_{M_{k_i}}, \sigma_{U_i}\in \mathbb{R}^+$; $\dot{W}_{k_i}(t)$ and $\dot{W}^\prime_i(t)$ are independent Gaussian white noise processes with $\textsf{E}\{\dot{W}_{k_i}(t)\}=\textsf{E}\{\dot{W}^\prime_i(t)\}=0$ and covariance function $\textsf{E}\{\dot{W}_{k_i}(t)\dot{W}_{k_i}(t^\prime)\}=\textsf{E}\{\dot{W}^\prime_i(t)\dot{W}^\prime_i(t^\prime)\}=\delta(t-t^\prime)$, a delta function. We denote $\bs{M}_{{k_i} 0} =\{M_{k_i}(0), D^1M_{k_i}(0), \dots,D^{p-1}M_{k_i}(0)\}$ and $\bs{U}_{i0}=\{U_i(0), D^1U_i(0), \dots, D^{q-1}U_i(0)\}$ as the initial values of $M_{k_i}(t)$ and $U_i(t)$ as well as their derivatives till orders $q-1$ and $p-1$ respectively. The volatility $\sigma^2_{U_i}$ in SDE \eqref{eq:sde_U_i} is allowed to vary between
subjects and across covariates. In this article, we focus on a simple transformed mean relationship, namely $\log(\sigma^2_{U_i}) \sim \mathsf{N}_1 (\bs{x}^\prime_i\bs{\beta},\sigma^2)$, which can be extended to the more complex specifications with less restrictive assumptions and to high-dimensional covariates.

The mean and covariance functions of Gaussian process priors for $M_{k_i}(t)$ and $U_{i}(t)$ can be obtained by applying stochastic integration to SDEs \eqref{eq:sde_M_k} and \eqref{eq:sde_U_i}, resulting in the following lemma.
\begin{lem}
\label{lem:GPs_nGP}
$M_{k}(t)$, $k = 1, 2, \dots, g$,  and $U_{i}(t)$, $i=1,2,\dots,n$, are the summations of mutually independent Gaussian processes written as $M_{k}(t) = M_{k0}(t) + M_{k1}(t)$ and $U_{i}(t) = U_{i0}(t) + U_{i1}(t)$ with corresponding mean functions $\textsf{E}\left\{M_{k0}(t)\right\}=\textsf{E}\left\{M_{k1}(t)\right\}=
 \textsf{E}\left\{U_{i0}(t)\right\}=\textsf{E}\left\{U_{i1}(t)\right\}=0$ and covariance functions 
\begin{align*}
\mathcal{K}_{M_{k0}}(s,t)&=\sigma^2_{M_0}\mathcal{R}_{M_{0}}(s,t)=\sigma^2_{M_0}\sum_{l=0}^{q-1}\phi_l(s)\phi_l(t),\\
\mathcal{K}_{M_{k1}}(s,t)&=\sigma^2_{M_k}\mathcal{R}_{M_{1}}(s,t)=\sigma^2_{M_k}\int_{\mathcal{T}}G_q(s,u)G_q(t,u)du,\\
\mathcal{K}_{U_{i0}}(s,t)&=\sigma^2_{U_0}\mathcal{R}_{U_{0}}(s,t)=\sigma^2_{U_0}\sum_{l=0}^{p-1}\phi_{l}(s)\phi_{l}(t),\\
\mathcal{K}_{U_{i1}}(s,t)&=\sigma^2_{U_i}\mathcal{R}_{U_{1}}(s,t)=\sigma^2_{U_i}\int_{\mathcal{T}}G_{p}(s,u)G_{p}(t,u)du,
\end{align*}
respectively, where $\phi_l(t)=\frac{t^l}{l!}$, $G_q(s,u)=\frac{(s-u)_{+}^{q-1}}{(q-1)!}$ and $s,t,u \in \mathcal{T}=[0,t_U]$.
\end{lem}
Hence, we can represent the prior of $M_{k_i}(t) + U_{i}(t)$ as a hierarchical Gaussian process, 
\begin{align*}
M_{k_i}(t) + U_{i}(t) \mid M_{k_i}(t) &\sim \mathcal{GP} (M_{k_i}(t), \mathcal{K}_{U_{i0}}(s,t)+\mathcal{K}_{U_{i1}}(s,t)), \\
M_{k_i}(t) &\sim \mathcal{GP} (0, \mathcal{K}_{M_{k0}}(s,t)+\mathcal{K}_{M_{k1}}(s,t)),
\end{align*}  
where $\mathcal{GP} (M(t), \mathcal{K}(s,t))$ denotes a Gaussian process with mean function $M(t)$ and covariance function $\mathcal{K}(s,t)$. Different from the previous hierarchical Gaussian process prior \citep{park2010hierarchical}, in which the covariance function is modeled as a squared exponential kernel and is identical across the subjects within a group, here $\mathcal{K}_{U_{i0}}(s,t)+\mathcal{K}_{U_{i1}}(s,t)$ is subject specific and depends on covariates through $\sigma^2_{U_i}$.
 
To carry out Bayesian inference, we further specify the following prior distributions for the hyperparameters. In particular,   
$\bs{M}_{{k_i} 0} \sim \mathsf{N}_{p} (\bs{0}, \sigma^2_{M_0} \bs{I})$ with $\sigma^2_{M_0}=10^4$,
$\bs{U}_{i0} \sim \mathsf{N}_{q} (\bs{0}, \sigma^2_{U_0} \bs{I})$, $\sigma^2_\varepsilon \sim \textsf{invGamma}(a,b)$, $\sigma^2_{M_k} \sim \textsf{invGamma}(a,b)$ and $\sigma^2_{U_0} \sim \textsf{invGamma}(a,b)$, where $\textsf{invGamma}(a,b)$ denotes the inverse gamma distribution with shape parameter $a$ and scale parameter $b$. The $\bs{\beta}$ and $\sigma^2$ follow the independent Jeffreys' prior, $f(\bs{\beta},\sigma^2) \propto 1/\sigma^2$.

\subsection{Double-Penalized Smoothing Spline}
It is well known that the Bayes estimate with the integrated Wiener process prior is identical to the smoothing spline estimate \citep{wahba1990spline}. By similar arguments, we can show that when the volatilities are given and $\sigma^2_{M_0}$ and $\sigma^2_{U_0}$ go to infinity, the posterior means of $M_{k}(t)$ and $U_{i}(t)$ are equivalent to the double-penalized smoothing spline $\hat{M}_{k}(t) + \hat{U}_{i}(t)$, which is the minimizer of the double-penalized sum-of-squares,
\begin{align}
\label{eq:DPSS}
\textsf{DPSS}=
&\sum_{i=1}^m\frac{1}{n_i}\sum_{j=1}^{n_i}\left\{Y(t_{ij})-M_{k_i}(t_{ij})-U_i(t_{ij})\right\}^2 +  \\ \notag
&\sum_{k=1}^g \lambda_{M_k}\int_{\mathcal{T}}\left\{D^{p}M_k(t)\right\}^2dt 
+\sum_{i=1}^m\lambda_{U_i}\int_{\mathcal{T}}\left\{D^{q}U_i(t)\right\}^2dt,
\end{align}
where penalty terms $\int_{\mathcal{T}}\left\{D^{p}M_k(t)\right\}^2dt$ and $\int_{\mathcal{T}}\left\{D^{q}U_i(t)\right\}^2dt$  penalize the roughness of $M_k(t)$ and $U_i(t)$ respectively, where the smoothness and the fidelity to data are balanced by the smoothing parameters $\lambda_{M_k}=\displaystyle\sum_{i:k_i=k}\frac{\sigma^2_\varepsilon}{n_i\sigma^2_{M_k}}$ and $\lambda_{U_i}=\frac{\sigma^2_\varepsilon}{n_i\sigma^2_{U_i}}$. Expression for $\hat{M}_{k}(t)$ and $\hat{U}_{i}(t)$ can be obtained explicitly, as shown in the following Theorem.   
\begin{thm}
\label{thm:DPSS}
The smoothing splines $\hat{M}_{k}(t)$ and $\hat{U}_{i}(t)$ with $t \in \mathcal{T}$ minimize the double-penalized sum-of-squares \eqref{eq:DPSS} and  have the forms
\begin{align*}
\hat{M}_{k}(t) &= 
\sum_{l=0}^{p-1} \mu_{kl}\phi_l(t) + \sum_{j=1}^n\nu_{kj}\mathcal{R}_{M_{1}}(t_j, t)
= 
\bs{\mu}^\prime_k\bs{\phi}_{\mu}(t) +
\bs{\nu}^\prime_k \bs{R}_{M_{1}}(t)  \\
\hat{U}_{i}(t) &= 
\sum_{l=0}^{q-1} \alpha_{il}\phi_l(t) +  \sum_{j=1}^{n_i}\gamma_{ij}\mathcal{R}_{U_{1}}(t_{ij}, t) 
=
\bs{\alpha}^\prime_i\bs{\phi}_\alpha(t) + 
\bs{\gamma}^\prime_i\bs{R}_{U_{i1}}(t)
\end{align*}
where
$\bs{\mu}_k=\{\mu_{k0},\mu_{k1},\cdots,\mu_{k(p-1)}\}^\prime$, 
$\bs{\nu}_k=(\nu_{k1},\nu_{k2},\cdots,\nu_{kn})^\prime$, $\bs{\alpha}_i=\left\{\alpha_{i0},\alpha_{i1},\cdots,\alpha_{i(q-1)}\right\}^\prime$ and
$\bs{\gamma}_i=(\gamma_{i1},\gamma_{i2},\cdots,\gamma_{in_i})^\prime$
are the coefficients for the bases
\begin{align*}
\bs{\phi}_{\mu}(t)&=\{\phi_0(t),\phi_1(t),\cdots,\phi_{p-1}(t)\}^\prime,\quad 
\bs{R}_{M_{1}}(t)=\{\mathcal{R}_{M_{1}}(t_1, t),\mathcal{R}_{M_{1}}(t_2, t),\cdots,\mathcal{R}_{M_{1}}(t_n, t)\}^\prime,\\
\bs{\phi}_\alpha(t)&=\{\phi_0(t),\phi_{1}(t),\cdots,\phi_{q-1}(t)\}^\prime,\quad
\bs{R}_{U_{i1}}(t)=\{\mathcal{R}_{U_{1}}(t_{i1}, t),\mathcal{R}_{U_{1}}(t_{i2}, t),\cdots,\mathcal{R}_{U_{1}}(t_{in_i}, t)\}^\prime,
\end{align*}
with $t_j \in \mathcal{T}_m = \cup_{i=1}^m \mathcal{T}_i = \{t_j,\; j= 1,2,\dots,n\}$, the index set of unique observation times among all $m$ subjects.

%\begin{cor}
%\label{cor:dpss1}
Given $\hat{M}_{k}(t)$ and $\hat{U}_{i}(t)$, the double-penalized sum-of-squares \eqref{eq:DPSS} can be written as
\begin{align}
\textsf{DPSS}=
&\sum_{i=1}^m\frac{1}{n_i} (\bs{Y}_i- \bs{\Delta}_i \bs{\phi}_\mu\bs{\mu}_{k_i} - \bs{\Delta}_i\bs{R}_{M_{1}}\bs{\nu}_{k_i} - \bs{\phi}_{\alpha_i}\bs{\alpha}_i-\bs{R}_{U_{i1}}\bs{\gamma}_i )^\prime \times \label{eq:DPSS2} \\ 
&(\bs{Y}_i- \bs{\Delta}_i \bs{\phi}_\mu\bs{\mu}_{k_i} - \bs{\Delta}_i\bs{R}_{M_{1}}\bs{\nu}_{k_i} - \bs{\phi}_{\alpha_i}\bs{\alpha}_i-\bs{R}_{U_{i1}}\bs{\gamma}_i )
 + \notag
\\ 
&\sum_{k=1}^g \lambda_{M_k}\bs{\nu}^\prime_{k} \bs{R}_{M_{1}}\bs{\nu}_{k}
+\sum_{i=1}^m\lambda_{U_i}\bs{\gamma}^\prime_i\bs{R}_{U_{i1}}\bs{\gamma}_i, \notag
\end{align}
where
\begin{align*} 
\bs{Y}_i&=\{Y(t_{i1}),Y(t_{i2}),\cdots,Y(t_{in_i})\}^\prime, 
\quad 
\bs{\Delta}_i = (\delta_{jj^\prime})_{n_i \times n},\\
\bs{\phi}_\mu&=\{\bs{\phi}_\mu(t_1),\bs{\phi}_\mu(t_2),\cdots,\bs{\phi}_\mu(t_n)\}^\prime, \quad
\bs{R}_{M_{1}} = \{\bs{R}_{M_{1}}(t_1), \bs{R}_{M_{1}}(t_2), \cdots, \bs{R}_{M_{1}}(t_n)\}, 
\\
\bs{\phi}_{\alpha_i}&=\{\bs{\phi}_\alpha(t_{i1}),\bs{\phi}_\alpha(t_{i2}),\cdots,\bs{\phi}_\alpha(t_{in_i})\}^\prime,
\quad
\bs{R}_{U_{i1}} =\{\bs{R}_{U_{i1}}(t_{i1}),\bs{R}_{U_{i1}}(t_{i2}),\cdots,\bs{R}_{U_{i1}}(t_{in_i})\}
\end{align*}
with $\delta_{jj^\prime} = 1$ if $i$th subject has an observation at time $t_{ij} = t_{j^\prime}$, $t_{ij} \in \mathcal{T}_i$,  $t_{j^\prime} \in \mathcal{T}_m$  and $\delta_{jj^\prime} = 0$, otherwise.  
%\end{cor}
\end{thm}
The proofs of Theorem \ref{thm:DPSS} and the following Corollary are included in Appendix \ref{sec:appendix_a}.
\begin{cor}
\label{cor:dpss2}
The $\bs{\mu}_k$, $\bs{\nu}_k$, $\bs{\alpha}_i$ and $\bs{\gamma}_i$ can be obtained through a backfitting algorithm or the Gauss-Seidel method, iterating the following two steps until convergence:
\begin{enumerate}[]
\item (a) for each $i$, 
$\bs{\hat{\alpha}}_i= (\bs{\phi}_{\alpha_i}^\prime \bs{S}_{U_i}^{-1} \bs{\phi}_{\alpha_i})^{-1}\bs{\phi}_{\alpha_i}^\prime \bs{S}_{U_i}^{-1} \bs{\tilde{Y}}_i$ and 
$\bs{\hat{\gamma}}_i=\bs{S}_{U_i}^{-1}\left\{\bs{I} - \bs{\phi}_{\alpha_i}(\bs{\phi}_{\alpha_i}^\prime \bs{S}_{U_i}^{-1} \bs{\phi}_{\alpha_i})^{-1}\bs{\phi}_{\alpha_i}^\prime \bs{S}_{U_i}^{-1} \right\}\bs{\tilde{Y}}_i$, where $\bs{S}_{U_i}=\bs{R}_{U_i1}+n_i\lambda_{U_i}\bs{I}$ and $\bs{\tilde{Y}}_i = \bs{Y}_i- \bs{\Delta}_i \bs{\phi}_\mu\bs{\hat{\mu}}_{k_i} - \bs{\Delta}_i\bs{R}_{M_{1}}\bs{\hat{\nu}}_{k_i}$;
\item (b) for each $k$, $\bs{\hat{\mu}}_k=(\bs{\phi}_\mu^\prime\bs{\Delta}^\prime\bs{S}_{M_k}^{-1}\bs{\Delta}\bs{\phi}_\mu)^{-1}
\bs{\phi}_\mu^\prime\bs{\Delta}^\prime\bs{S}_{M_k}^{-1}\bs{\tilde{Y}}_k$ and \\
$\bs{\hat{\nu}}_k=\bs{S}_{M_k}^{-1}
\left\{
\bs{I}-\bs{\Delta}\bs{\phi}_\mu(\bs{\phi}_\mu^\prime\bs{\Delta}^\prime\bs{S}_{M_k}^{-1}\bs{\Delta}\bs{\phi}_\mu)^{-1}
\bs{\phi}_\mu^\prime\bs{\Delta}^\prime\bs{S}_{M_k}^{-1}
\right\}\bs{\tilde{Y}}_k$, where $\bs{S}_{M_k} = \bs{\Delta}\bs{R}_{M_1}+\lambda_{M_k}\bs{I}$, $\bs{\tilde{Y}}_k=\displaystyle\sum_{i:k_i=k}\frac{1}{n_i}\bs{\Delta}_i^\prime\left(\bs{Y}_i-  \bs{\phi}_{\alpha_i}\bs{\hat{\alpha}}_i-\bs{R}_{U_{i1}}\bs{\hat{\gamma}}_i \right )$ and $\bs{\Delta}= \displaystyle\sum_{i:k_i=k}\frac{1}{n_i}\bs{\Delta}_i^\prime\bs{\Delta}_i $.
\end{enumerate}
\end{cor}
  
\section{Posterior Computation}
\label{sec:posterior}
Although we can obtain $\hat{M}_k(t)$ and $\hat{U}_i(t)$ by the backfitting algorithm outlined in Corollary \ref{cor:dpss2} and estimate $\lambda_{M_k}$ and $\lambda_{U_i}$ through generalized cross validation \citep[Chap. 4, ][]{wahba1990spline}, it is unclear how to allow $\lambda_{U_i}$ to depend on covariates. In addition,  when $n$ is large,  it is computational infeasible to invert the $n \times n$ matrix $\bs{S}_{M_k}$ involved in the backfitting algorithm. Instead, we propose a Markov chain Monte Carlo (MCMC) algorithm for posterior computation that solves these problems. The algorithm achieves computational efficiency by leveraging on the Markovian property of SDEs and samples $M_k(t)$ and $U_i(t)$ through the simulation smoother \citep{durbin2002simple}, which requires the following Proposition.

\begin{prop}
\label{prop:IW}
Let $X(t)$ denote a (r-1)th-order integral Wiener process,  defined by the stochastic differential equation 
$D^r X(t)=\dot{W}(t)$. Consequently, the $\bs{X}_j=\{X(t_j), D^1X(t_j),\cdots,D^{r-1}X(t_j)\}^\prime$, $j=1,2,\cdots,n$, follows a state equation 
\begin{equation*}
\bs{X}_{j+1}=\bs{G}_j\bs{X}_j+\bs{\omega}_j,
\end{equation*}
where $\bs{G}_j=\sum_{k=0}^r\frac{\delta_j^k}{k!}\bs{C}^k$ and $\bs{\omega}_j \sim \mathsf{N}_r(\bs{0},\bs{W}_j)$ with $\bs{W}_j=\int_0^{\delta_j}\exp\{\bs{C}(\delta_j-u)\}\bs{D}\bs{D}^\prime\exp\{\bs{C}^\prime(\delta_j-u)\}du$,
$\bs{C}=(c_{ll^\prime})_{r \times r}$, $c_{ll^\prime}=1$ when $l^\prime=l+1$ and $c_{ll^\prime}=0$ otherwise, $\bs{D}=(0,0,\cdots,1)^\prime$ and $\delta_j=t_{j+1}-t_j$. 
\end{prop}
\noindent The proof is in Appendix \ref{sec:appendix_a}. Finally, we outline the proposed MCMC as follows.
\begin{enumerate}[]
\item (1) Given $M_{k_i}(t_{ij})$, $\sigma^2_\varepsilon$ and $\sigma^2_{U_i}$, sample $U_i(t_{ij})$, $i = 1, 2, \cdots, m$, $j=0,1,\cdots,n_i$. Let $Y_{U_{ij}}=Y_i(t_{ij})-M_{k_i}(t_{ij})$ and the SVR model for the $i$th subject can be expressed as the following state space model \citep{jones1993longitudinal,durbin2001time}, from which we can draw samples of $U_i(t_{ij})$ and its derivatives using the simulation smoother.  
\begin{align*}
Y_{U_{ij}} &= \bs{F}_{U_{ij}}\bs{U}_{{ij}} + \varepsilon_{U_{ij}},\\
\bs{U}_{{i(j+1)}} &= \bs{G}_{U_{ij}}\bs{U}_{{ij}}+\sigma_{U_i}\bs{\omega}_{U_{ij}},
\end{align*}
where $\bs{F}_{U_{ij}}=(1,0,\cdots,0)$, $\bs{U}_{{ij}}=\{U_i(t_{ij}), D^1U_i(t_{ij}),\cdots,D^{q-1}U_i(t_{ij})\}^\prime$ and $\varepsilon_{U_{ij}} \iid \mathsf{N}_1(0, \sigma^2_\varepsilon)$. Similar to the $\bs{G}_{j}$, $\bs{\omega}_{j}$ and $\bs{W}_{j}$ in Proposition \ref{prop:IW},  the $\bs{G}_{U_{ij}}$, $\bs{\omega}_{U_{ij}}$ and $\bs{W}_{U_{ij}}$ follow the same specifications with $r=q$.
\item (2) Given $U_i(t_j)$, $\sigma^2_\varepsilon$ and $\sigma^2_{M_k}$, sample $M_k(t_j)$, $k=1,2,\cdots,g$, $j=0,1,\cdots,n$. Similarly, we rewrite the SVR model for the $k$th group as the following state space model and then sample $M_{k_i}(t_{ij})$ and its derivatives by the simulation smoother. 
\begin{align*}
\bs{Y}_{M_{kj}}&=\bs{F}_{M_{kj}}\bs{M}_{kj}+\bs{\varepsilon}_{M_{kj}},\\
\bs{M}_{{k(j+1)}} &= \bs{G}_{M_{kj}}\bs{M}_{{kj}}+{\sigma}_{M_{k}}\bs{\omega}_{M_{kj}},
\end{align*}
where $\bs{Y}_{M_{kj}}=(Y_{M_{kj}}^i)_{m \times 1}$, $\bs{M}_{{kj}}=\{M_k(t_{j}), D^1M_k(t_{j}),\cdots,D^{p-1}M_k(t_{j})\}^\prime$, $\bs{F}_{M_{kj}}=(F_{M_{kj}}^{il})_{m\times p}$ and $\bs{\varepsilon}_{M_{kj}}=\text{diag}(\varepsilon_{M_{kj}}^1,\varepsilon_{M_{kj}}^2,
\cdots,\varepsilon_{M_{kj}}^m)$. When $i$th subject has an observation at time $t_j$ and $k_i=k$, $Y_{M_{kj}}^i=Y_i(t_j)-U_i(t_j)$, $F_{M_{kj}}^{i1}=1$ and $\varepsilon_{M_{kj}}^i \sim \mathsf{N}_1(0, \sigma^2_\varepsilon)$. Otherwise, $Y_{M_{kj}}^i=F_{M_{kj}}^{il}=\varepsilon_{M_{kj}}^i=0$. The $\bs{G}_{M_{kj}}$, $\bs{\omega}_{M_{kj}}$ and $\bs{W}_{M_{kj}}$ are given by Proposition \ref{prop:IW} with $r=p$.
\item (3a) Given $M_{k_i}(t_{ij})$ and $U_i(t_{ij})$, $i=1,2,\cdots,m$, $j=1,2,\cdots,n_i$, sample $\sigma^2_\varepsilon$ from the posterior distribution $\textsf{invGamma}\left(a+\frac{1}{2}\sum_{i=1}^m n_i,b+\frac{1}{2}\sum_{i=1}^m\sum_{j=1}^{n_i}\left
\{Y_i(t_{ij})-M_{k_i}(t_{ij})-U_i(t_{ij})\right\}^2\right)$.

\item (3b) Given $\bs{U}_{{i0}}$, sample $\sigma^2_{U_0}$ from the posterior distribution 
 $\textsf{invGamma}\left(a+\frac{mq}{2},b+\frac{1}{2}\sum_{i=0}^{m}
\bs{U}_{{i0}}^\prime\bs{U}_{{i0}}\right)$.

\item (3c) Given $\bs{M}_{{kj}}$, sample $\sigma_{M_k}^2$ from the posterior distribution \\ $\textsf{invGamma}\left(a+\frac{np}{2},b+\frac{1}{2}\sum_{j=0}^{n-1}
(\bs{M}_{{k(j+1)}}- \bs{G}_{M_{kj}}\bs{M}_{{kj}})^\prime\bs{W}_{M_{kj}}^{-1}(\bs{M}_{{k(j+1)}}- \bs{G}_{M_{kj}}\bs{M}_{{kj}})\right)$.

\item (3d) Given $\bs{U}_{{ij}}$, $\bs{\beta}$ and $\sigma^2$, sample $\sigma^{2}_{U_i}$ using a Metropolis-Hasting algorithm. We choose $\sigma^{2}_{U_i} \sim \textsf{invGamma}(a, b)$ as the proposal prior distribution and a proposal $\sigma^{2*}_{U_i}$ can be easily drawn from $\textsf{invGamma}\left(a+\frac{n_iq}{2},b+\frac{1}{2}\sum_{j=0}^{n_i-1}
(\bs{U}_{{i(j+1)}}- \bs{G}_{U_{ij}}\bs{U}_{{ij}})^\prime\bs{W}_{U_{ij}}^{-1}(\bs{U}_{{i(j+1)}}- \bs{G}_{U_{ij}}\bs{U}_{{ij}})\right)$ the corresponding proposal posterior distribution. The  $\sigma^{2*}_{U_i}$ will be accepted with the following probability and discarded otherwise with $\sigma^{2}_{U_i}$ unchanged, 
\begin{equation*}\min \left\{
\frac{
f_{\mathsf{LN}}(\sigma^{2*}_{U_i} \mid \bs{x}^\prime_i\bs{\beta},\sigma^2)
\prod_{j=0}^{n_i-1}
f_{\mathsf{N_q}}( \bs{U}_{{i(j+1)}}- \bs{G}_{U_{ij}}\bs{U}_{{ij}} \mid \bs{0},\sigma^{2*}_{U_i}\bs{W}_{U_{ij}})
f_{\mathsf{iG}}(\sigma^{2}_{U_i} \mid a_{U_i},b_{U_i})
}{
f_{\mathsf{LN}}(\sigma^{2}_{U_i} \mid \bs{x}^\prime_i\bs{\beta},\sigma^2)
\prod_{j=0}^{n_i-1}
f_{\mathsf{N_q}}( \bs{U}_{{i(j+1)}}- \bs{G}_{U_{ij}}\bs{U}_{{ij}} \mid \bs{0},\sigma^{2}_{U_i}\bs{W}_{U_{ij}})
f_{\mathsf{iG}}(\sigma^{2*}_{U_i} \mid a_{U_i},b_{U_i})
}
,1
\right\},
\end{equation*}
where $f_{\mathsf{LN}}$, $f_{\mathsf{N_q}}$ and $f_{\mathsf{iG}}$ denote the log-normal, $q$-dimensional normal and inverse gamma probability density functions respectively with $a_{U_i}=a+\frac{n_iq}{2}$, $b_{U_i}=b+\frac{1}{2}\sum_{j=0}^{n_i-1}
(\bs{U}_{{i(j+1)}}- \bs{G}_{U_{ij}}\bs{U}_{{ij}})^\prime\bs{W}_{U_{ij}}^{-1}(\bs{U}_{{i(j+1)}}- \bs{G}_{U_{ij}}\bs{U}_{{ij}})$.
\item (4) Given $\sigma^2_{U_i}$, sample $\bs{\beta}$ and $\sigma^2$.
Let $\bs{Z}=(\log\sigma^2_{U_1}, \log\sigma^2_{U_2},\cdots, \log\sigma^2_{U_m})^\prime$, $\bs{\hat{\beta}}=(\bs{X}^\prime\bs{X})^{-1}\bs{X}^\prime\bs{Z}$ and $\hat{\sigma}^2=\frac{(\bs{Z}-\bs{X}\bs{\hat{\beta}})^\prime(\bs{Z}-\bs{X}\bs{\hat{\beta}})}{m-k}$.
We draw $\tau$ from Chi-squared distribution with $m-k$ degrees of freedom and set $\sigma^2=\frac{(m-k)\hat{\sigma}^2}{\tau}$ and then sample $\bs{\beta}$ from $\textsf{N}_m \left( \bs{\hat{\beta}}, \sigma^2(\bs{X}^\prime\bs{X})^{-1}
\right)$.

\end{enumerate}

\section{Simulation}
\label{sec:simu}
We carry out two simulation studies to evaluate the performance of the proposed method and compare it to alternative methods including nature cubic spline \citep[NCS, ][]{wahba1990spline} and functional principal components analysis \citep[FPCA, ][]{yao2005functional}. The comparison focuses on performance in estimating the trajectory $M_{k_i}(t) + U_{i}(t)$, the volatility $\sigma^2_{U_i}$ and the coefficients $\bs{\beta}$. 

The first simulation study is designed to investigate the consequence of ignoring either similarity or heterogeneity of volatilities when they are present. One hundred replicated datasets, each consisting of 100 trajectories, are sampled from the SVR model, in which the log-transformed volatilities are varying and normally distributed. More precisely, we choose $\bs{\beta}=(0, 0.6, 2)^\prime$ and $\bs{x}_i=(1, x_{i1}, x_{i2})^\prime$ with $x_{i1}$ and $x_{i2}$ sampled from $x_{i1} \iid \mathsf{Bernoulli}(0.4)$ and $x_{i2} \iid \mathsf{N}_1(0, 0.25)$ respectively. Given $\bs{\beta}$ and $\bs{x}_i$, volatilities $\sigma^2_{U_i}$'s can be drawn from $\log(\sigma^2_{U_i}) \sim \mathsf{N}_1 (\bs{x}^\prime_i\bs{\beta},1)$. Along with $\sigma^2_{M_1}=\sigma^2_{M_2}=10$, $\sigma^2_\varepsilon =1$, $p=2$ and $q=1$, $M_1(t)$, $M_2(t)$, $U_i(t)$ and $\varepsilon_i(t)$ are sampled at $t \in \{0.2, 0.4, \cdots, 4\}$ from equations \eqref{eq:sde_M_k} and \eqref{eq:sde_U_i} and the distribution of measurement error $\varepsilon_i(t)$.  Twenty percent of samples are removed completely at random, resulting in on average 16 unequally spaced observations per subject. Finally, the $i$th subject is randomly assigned to one of the two groups with equal probability and $Y_i(t)$ is obtained from observation equation \eqref{eq:obs}. 

The simulated datasets are analyzed by three methods, SVR, NCS and FPCA. We first apply the SVR approach, using the proposed MCMC algorithm with 15,000 iterations and keeping every 5th of the last 10,000 samples for posterior analysis. It takes about 80 minutes on a PC with 2.33GHz Intel(R) Xeon(R) CPU. Posterior means are chosen as the estimates of $M_{k_i}(t) + U_i(t)$, $\sigma^2_{U_i}$ and $\bs{\beta}$.  Additionally, the trajectories $M_{k_i}(t) + U_i(t)$'s are estimated by NCS for one subject at a time, and by FPCA for all subjects within a group and separately by the group, taking about 1 minute and 2 minutes on the same PC respectively. For NCS and FPCA methods, we may also estimate covariate effects on volatility through a two-stage method: estimating empirical volatility by $\frac{1}{n_i}\sum_{j=1}^{n_i-1}\frac{( \hat{U}_{i,j+1}-\hat{U}_{i,j} )^2}{t_{i,j+1}-t_{i,j}}$ in the first stage with $\hat{U}_{i,j}$ the estimate of $U_i(t)$ at time $t_{ij}$; and in the second stage,  empirical volatilities are regressed on covariates to obtain the estimate of $\bs{\beta}$.

For each simulated dataset, we calculate average squared error (ASE) for the trajectory $\text{ASE}(M + U) = \frac{1}{m}\sum_{i=1}^{m}\frac{1}{n_i}\sum_{j=1}^{n_i}\left\{ \hat{M}_{k_i}(t_{ij}) + \hat{U}_i(t_{ij}) - M_{k_i}(t_{ij}) - U_i(t_{ij})  \right\}^2$, ASE for log volatility $\text{ASE}\{log(\sigma^2_U)\}=\frac{1}{m}\sum_{i=1}^{m}\left\{ log(\hat{\sigma}^2_{U_i}) - log(\sigma^2_{U_i})  \right\}^2$, and squared errors (SE) for coefficient estimates $\text{SE}(\beta_l) = (\hat{\beta}_l-\beta_l)^2$, $l=0,1,2$. Table \ref{tbl:MSE} reports means of $\text{ASE}(M + U)$,  $\text{ASE}\{log(\sigma^2_U)\}$ and $\text{SE}(\beta_l)$ across 100 replicate datasets. MASEs and MSEs by NCS and FPCA approaches are significantly inflated, for example, being doubled and tripled in $\text{MASE}(M + U)$ respectively, compared to SVR. We randomly select a data set and take a close examination. We calculate the individual ASE of the trajectory $\frac{1}{n_i}\sum_{j=1}^{n_i}\left\{ \hat{M}_{k_i}(t_{ij}) + \hat{U}_i(t_{ij}) - M_{k_i}(t_{ij}) - U_i(t_{ij})  \right\}^2$ and select the top four subjects with the largest individual ASEs with respect to NCS and FPCA approaches respectively. 

\begin{figure}
\centering
\subfloat[ID=17]{\label{fig:MU17}\includegraphics[width=0.22\textwidth,angle=270]{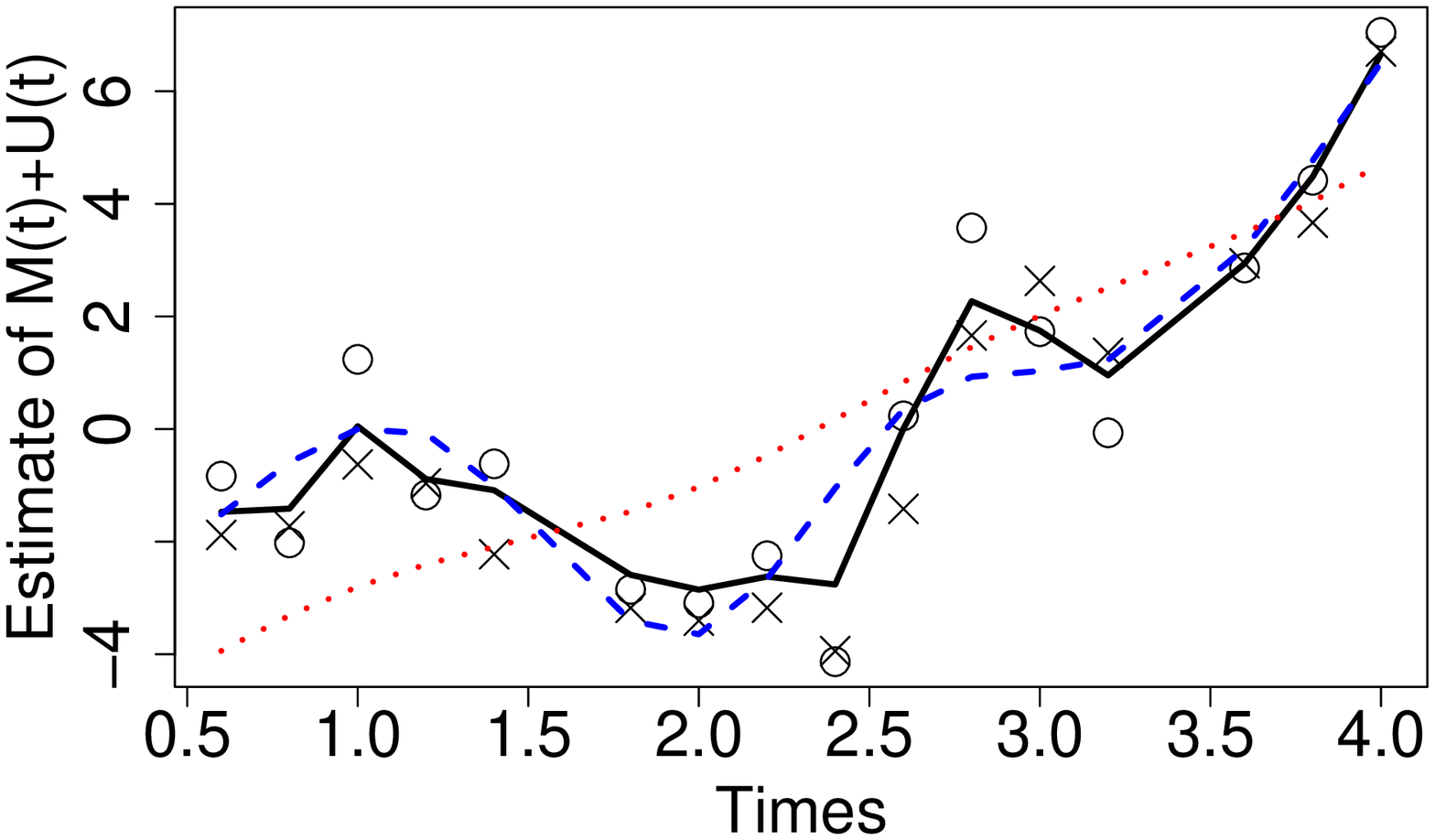}}
\subfloat[ID=30]{\label{fig:MU30}\includegraphics[width=0.22\textwidth,angle=270]{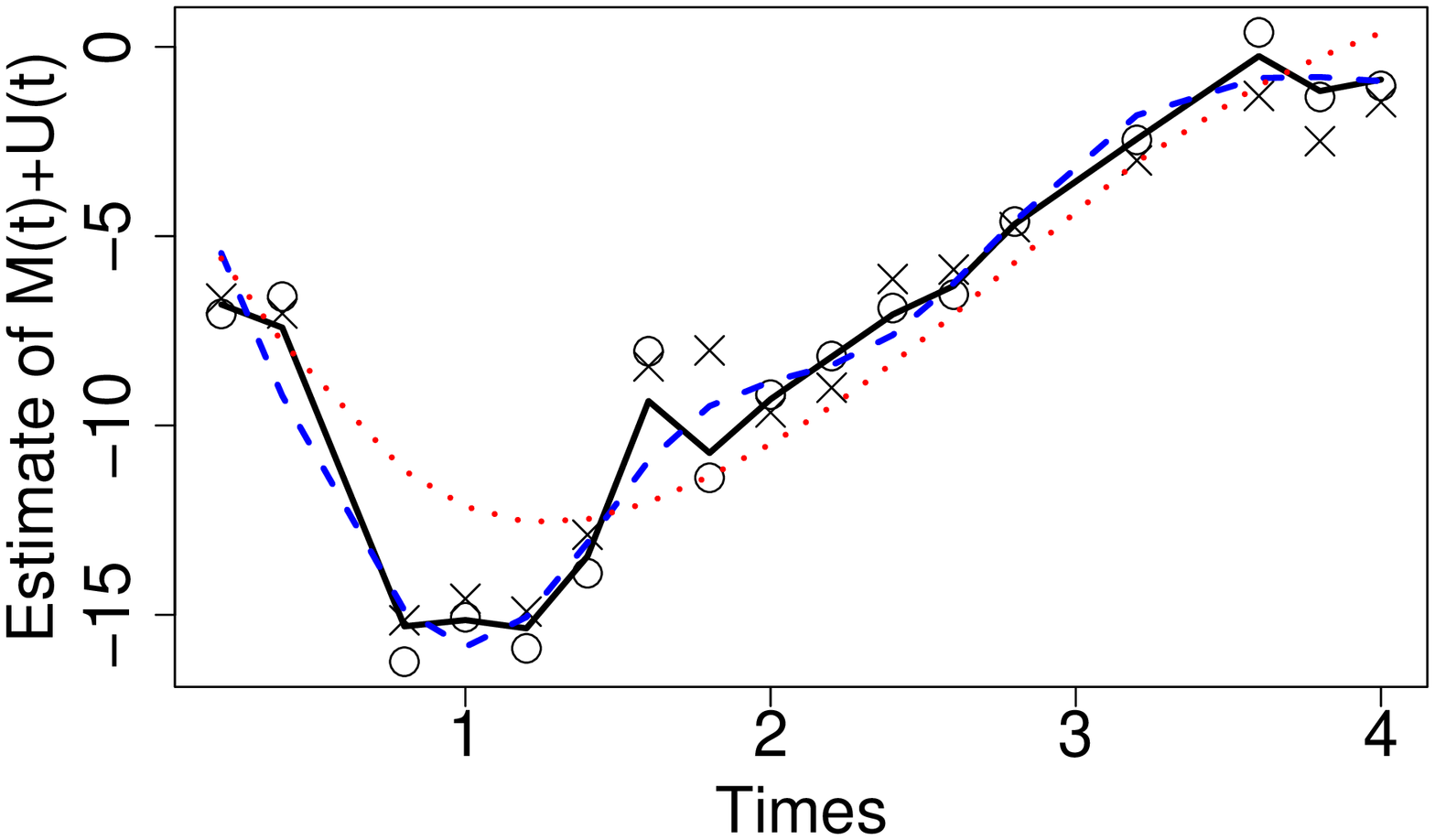}}
\subfloat[ID=31]{\label{fig:MU31}\includegraphics[width=0.22\textwidth,angle=270]{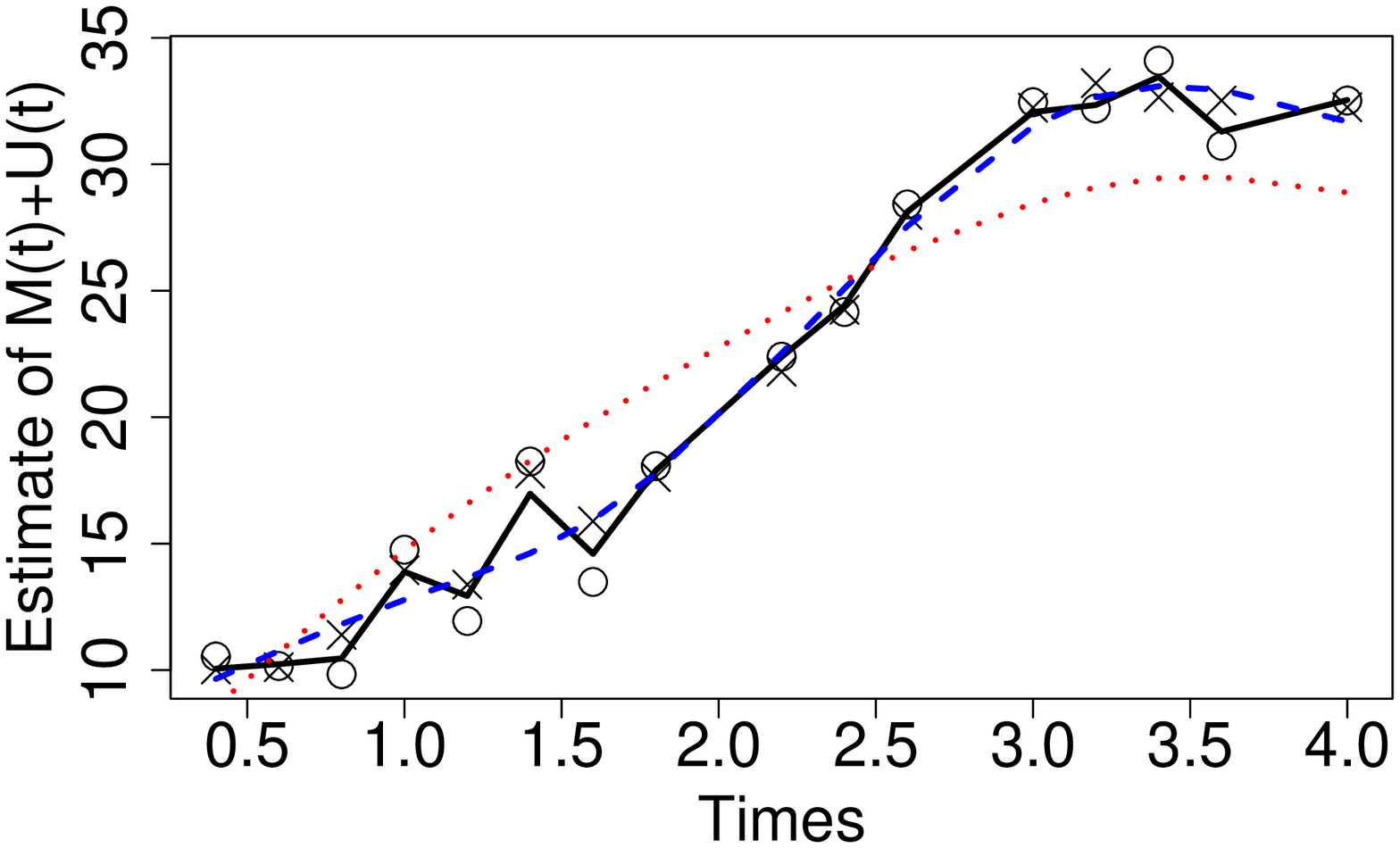}}\\
\subfloat[ID=39]{\label{fig:MU39}\includegraphics[width=0.22\textwidth,angle=270]{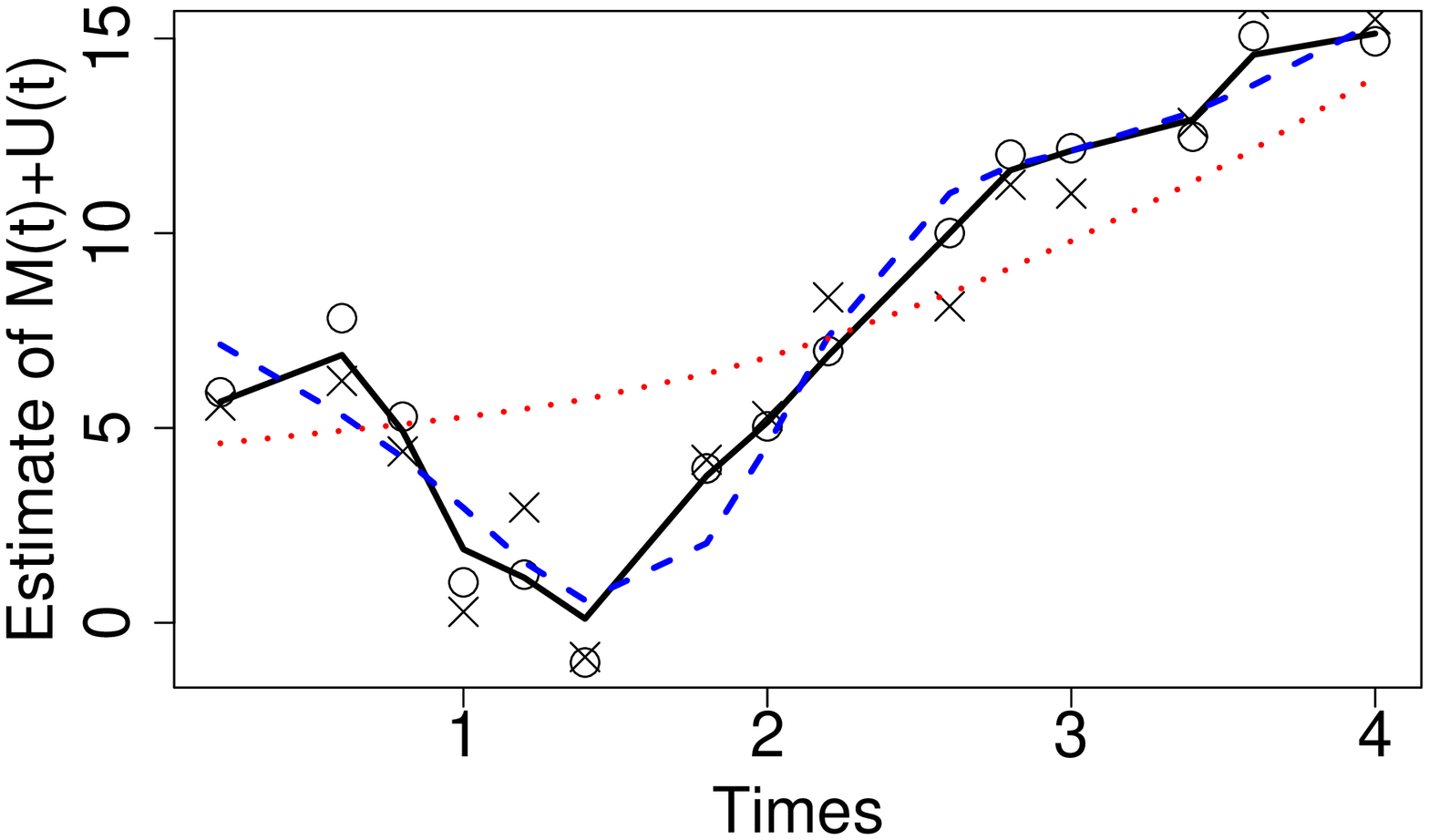}}
\subfloat[ID=44]{\label{fig:MU44}\includegraphics[width=0.22\textwidth,angle=270]{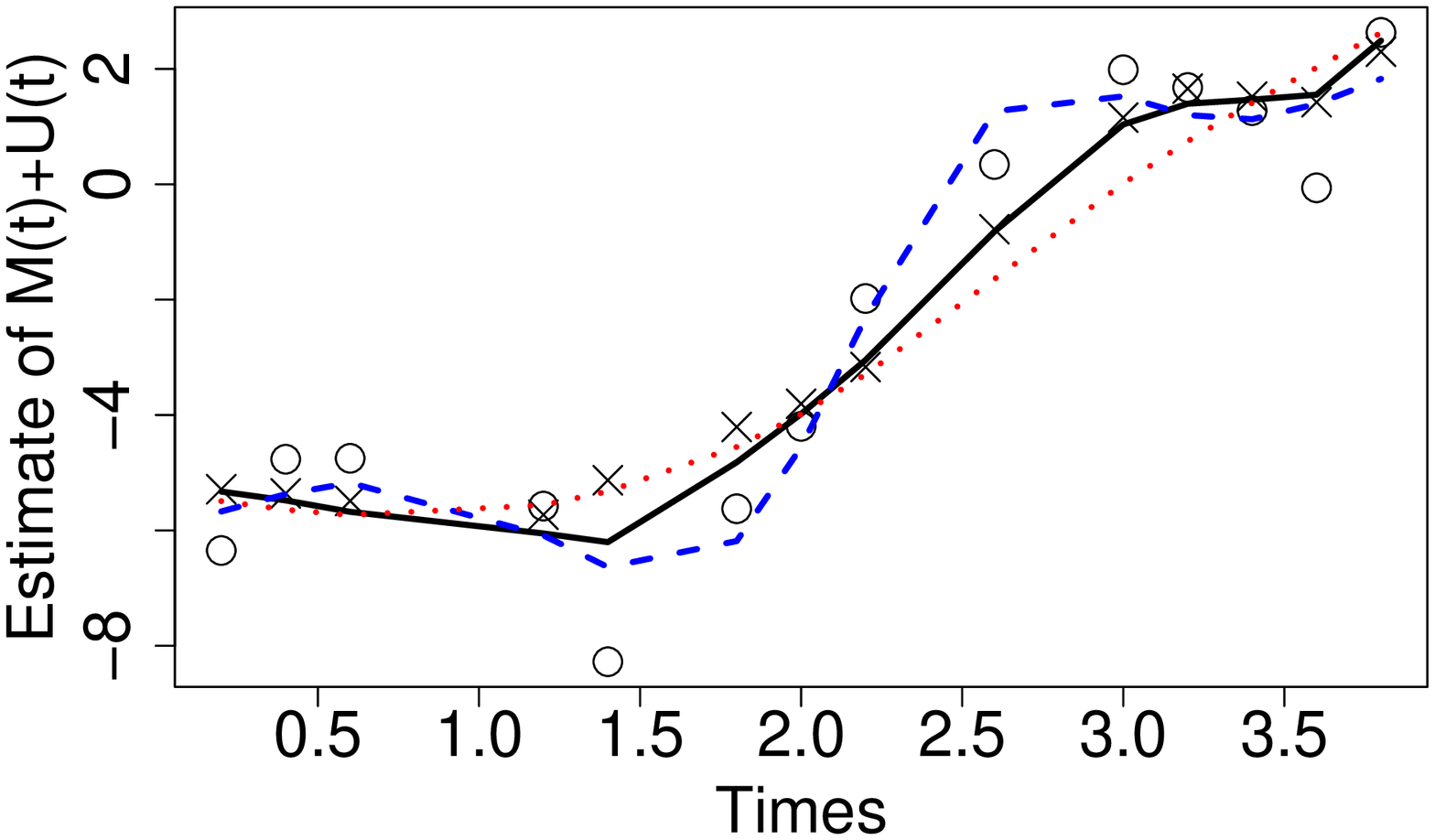}}
\subfloat[ID=86]{\label{fig:MU86}\includegraphics[width=0.22\textwidth,angle=270]{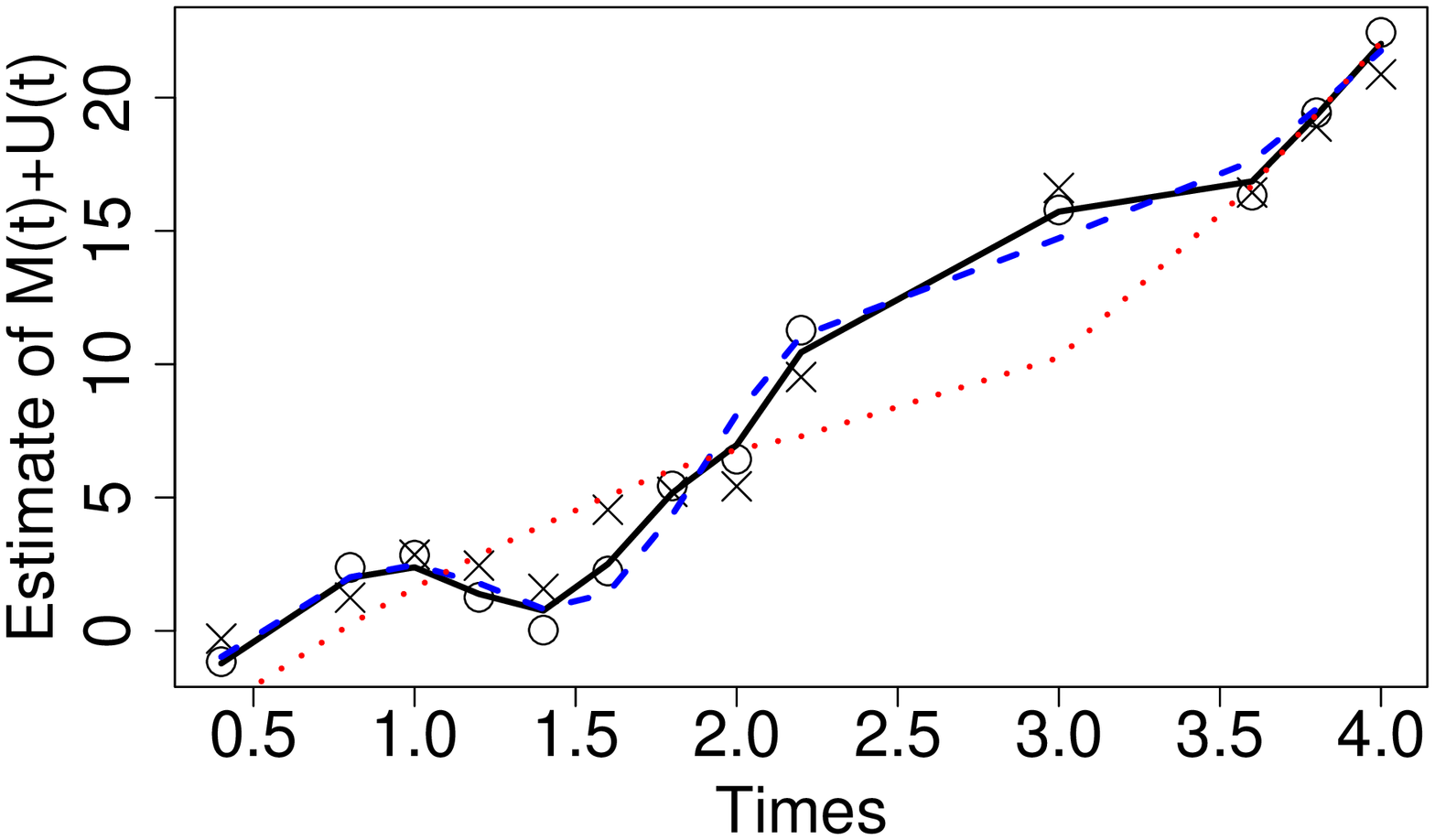}}

\caption{\label{fig:MU} The plots of observation ($\circ$) and  trajectory at time $t_{ij}$ ($\times$), as well as estimates of  trajectory $M_{k_i}(t)+U_i(t)$ by SVR (\textemdash), NCS ($---$) and  FPCA ($\cdotp \cdotp \cdotp$) approaches,  for six subjects in one simulated dataset with the largest individual average squared errors  $\frac{1}{n_i}\sum_{j=1}^{n_i}\left\{ \hat{M}_{k_i}(t_{ij}) + \hat{U}_i(t_{ij}) - M_{k_i}(t_{ij}) - U_i(t_{ij})  \right\}^2$.}    
\end{figure}

Figure \ref{fig:MU} shows estimates of the trajectory for six subjects.  The plots reveal that the increased MASEs or MSEs by NCS and FPCA are caused by different reasons. NCS approach, treating one trajectory a time, ignores the similarity between the subjects within a group, leading to over fitting true trajectories (e.g. Figure \ref{fig:MU}(d) and \ref{fig:MU}(e)) with both over and under estimated volatilities. On the other hand, FPCA approach omits the heterogeneity of the subjects within a group; inflated $\text{MASE}(M + U)$ are mainly contributed by a few subjects whose trajectory fluctuates with significantly higher volatility but is overly smoothed (e.g. Figure \ref{fig:MU}(b) and \ref{fig:MU}(d)); and under the assumption of similar smoothness, the estimates of volatility are largely under estimated. Although this simulation study is in favor of SVR approach by design, the scenario we consider is nevertheless realistic in practice and the simulation results reveal the drawbacks of omitting similarity or heterogeneity of volatilities by alternative approaches.    

Our second simulation study is conducted to evaluate the performance of SVR, NCS and FPCA when volatilities are homogeneous with no covariate effects. As in the first simulation study, 100 replicate datasets are generated, each consisting of 100 trajectories at $t \in \{0.2, 0.4, \cdots, 4\}$; twenty percent of data points are deleted completely at random; and subjects are assigned to one of two groups with equal probability. The observations are generated from $Y_i(t) = 10\{t+sin(t)\}+ 0.6\alpha_{1i}cos(\pi t/10)+0.2\alpha_{2i}sin(\pi t/10)+\varepsilon_{i}(t)$ for subjects in the first group and from $Y_i(t) = 10\{t+cos(t)\}+ 0.5\alpha_{1i}cos(\pi t/10)+0.3\alpha_{2i}sin(\pi t/10)+\varepsilon_{i}(t)$ for the ones in the second group, with $\alpha_{1i} \iid \mathsf{N}_1(0,4)$, $\alpha_{2i} \iid \mathsf{N}_1(0,1)$ and $\varepsilon_{i}(t) \iid \mathsf{N}_1(0, 1)$. The SVR, NCS and FPCA approaches are applied and $\text{MASE}(M + U)$ is presented in Table \ref{tbl:MSE}, in which SVR and FPCA approaches show close $\text{MASE}(M + U)$, both smaller than the one by NCS approach. This suggests that SVR is no worse than FPCA  for the cases with homogeneous volatilities.

In short, through the two simulation studies, we demonstrate that SVR achieves substantial gains in terms of reducing the ASEs or SEs of the estimates of the trajectory, volatility and covariate effect when volatilities are heterogeneous, and works at least as well as FPCA approach when volatilities are homogeneous.      

\begin{table}[ht]
\caption{The mean of squared errors or average square errors of the estimates of trajectory, volatility and covariate effect across 100 replicate datasets .\label{tbl:MSE}}
\vspace{5pt}
\begin{center}
\begin{tabular}{ccccccc}
  \hline\hline
&&&Case I&& & Case II \\
\cline{2-7} 
  method & $M+U$ & $\log(\sigma^2_U)$ & $\beta_0$ & $\beta_1$ & $\beta_2$ & $M+U$ \\ 
  \hline
 SVR & 0.345 &0.614&0.043&0.081&0.075&1.122\\
  NCS& 0.609 &1.297 &0.089&0.165&1.724&1.477\\
 FPCA&1.099 &2.966 &1.144&0.185&1.969&1.185\\
\hline 
\end{tabular}
\end{center}
\end{table}

%The convergence of the algorithm is indicated by trace plots and autocorrelation plots. Figure \ref{fig:Simu} plots the posterior means and $95\%$ credible intervals for the volatilities  $\sigma^2_{U_i}$'s and coefficients $\bs{\beta}$ respectively. It is clear that almost all of the credible intervals of volatilities and coefficients cover the true values and the posterior means are close to the true ones, both of which suggest the proposed MCMC algorithm work well. 
% 

%\begin{figure}
%\subfloat[]{\label{fig:SimuVol}\includegraphics[width=0.50\textwidth,angle=270]{Figures/SimuVolatility}}                 
%\subfloat[]{\label{fig:SimuCoeff}\includegraphics[width=0.50\textwidth,angle=270]{Figures/SimuCoeff}}  \caption{The posterior plots of the volatilities and coefficients versus the true values. The red dots are posterior means and bars are $95\%$ credible intervals. \label{fig:Simu}}
%\end{figure}
    
\section{Applications}
\label{sec:app}
It is a common practice to monitor the blood pressure of pregnant woman during pregnancy. Despite the trend of blood pressure being well studied, its fluctuation has been little addressed and the factors associated with the fluctuation are largely unknown. In this analysis, we apply the proposed SVR approach to analyze longitudinal blood pressure measurements in HPHB study, aiming to investigate the stability of blood pressure trajectories and identify the associated factors. The data set consists of 106 non-Hispanic white (NHW) and 176 non-Hispanic black (NHB) women whose first blood pressure measurement is collected before the 16th week of gestation and the last one no earlier than the 37th week of gestation. Most of subjects have 9 ($35.10\%$ of them), 10 ($29.28\%$) or 11 ($14.98\%$) measurements spaced at irregularly times. The covariates we focused on include race as NHW vs NHB, mother's age group as age $>$ 35 vs age $\le$ 35, obesity as yes vs no, preeclampsia as yes vs no, parity group as parity $>$ 0 vs parity $=$ 0, and smoking as yes vs no. We run the proposed MCMC algorithm for 15,000 iterations, discard the first 5,000 and retain every 5th of the remaining samples for posterior analysis.
The trace plots and autocorrelation plots suggest the algorithm converges fast and mixes well. Posterior summary of selected parameters is presented in Table \ref{tbl:post}.

\begin{figure}
\centering
\subfloat[]{\label{fig:MAPMNHW}\includegraphics[width=0.30\textwidth,angle=270]{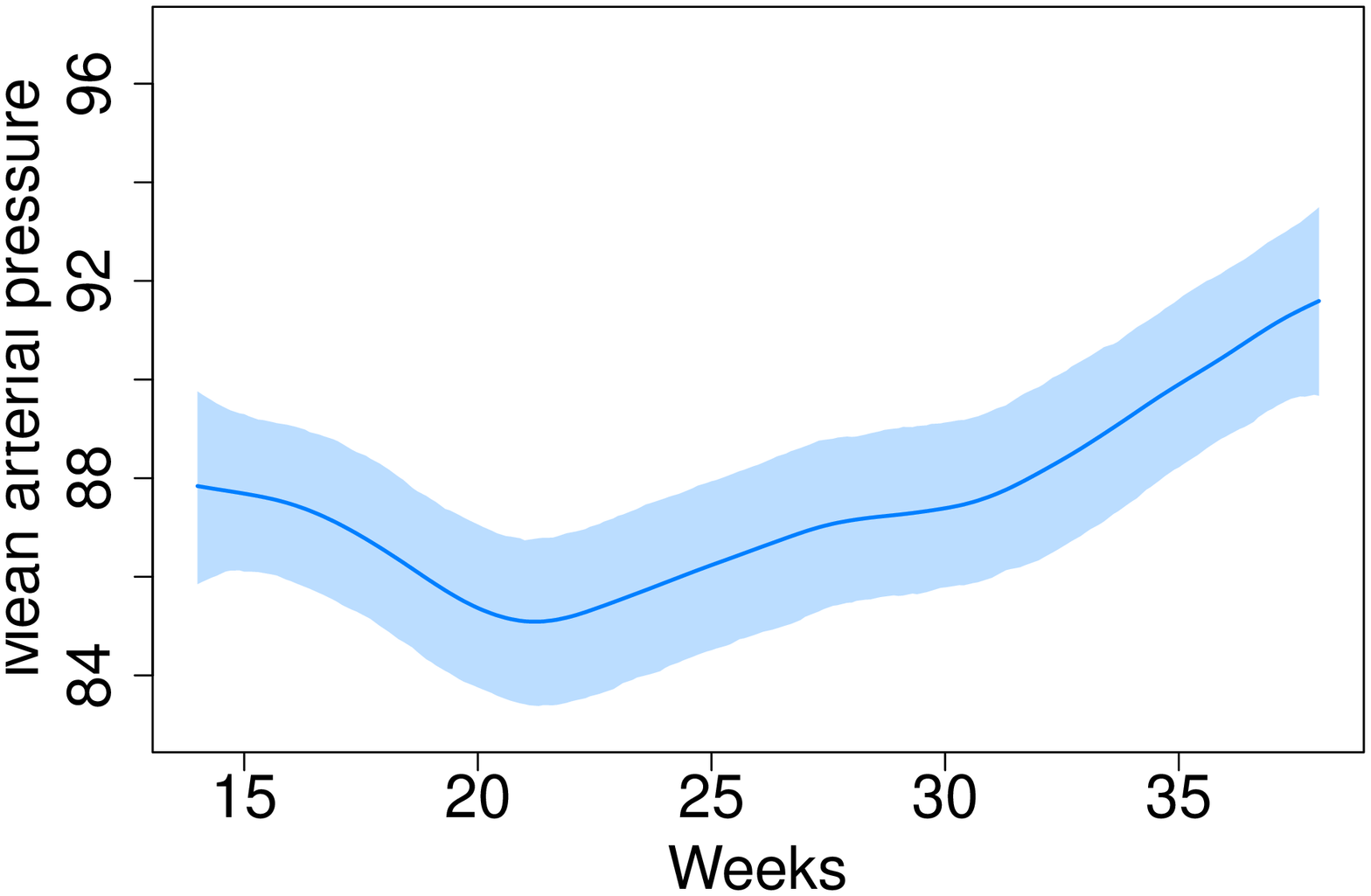}}
\qquad
\subfloat[]{\label{fig:MAPMNHB}\includegraphics[width=0.30\textwidth,angle=270]{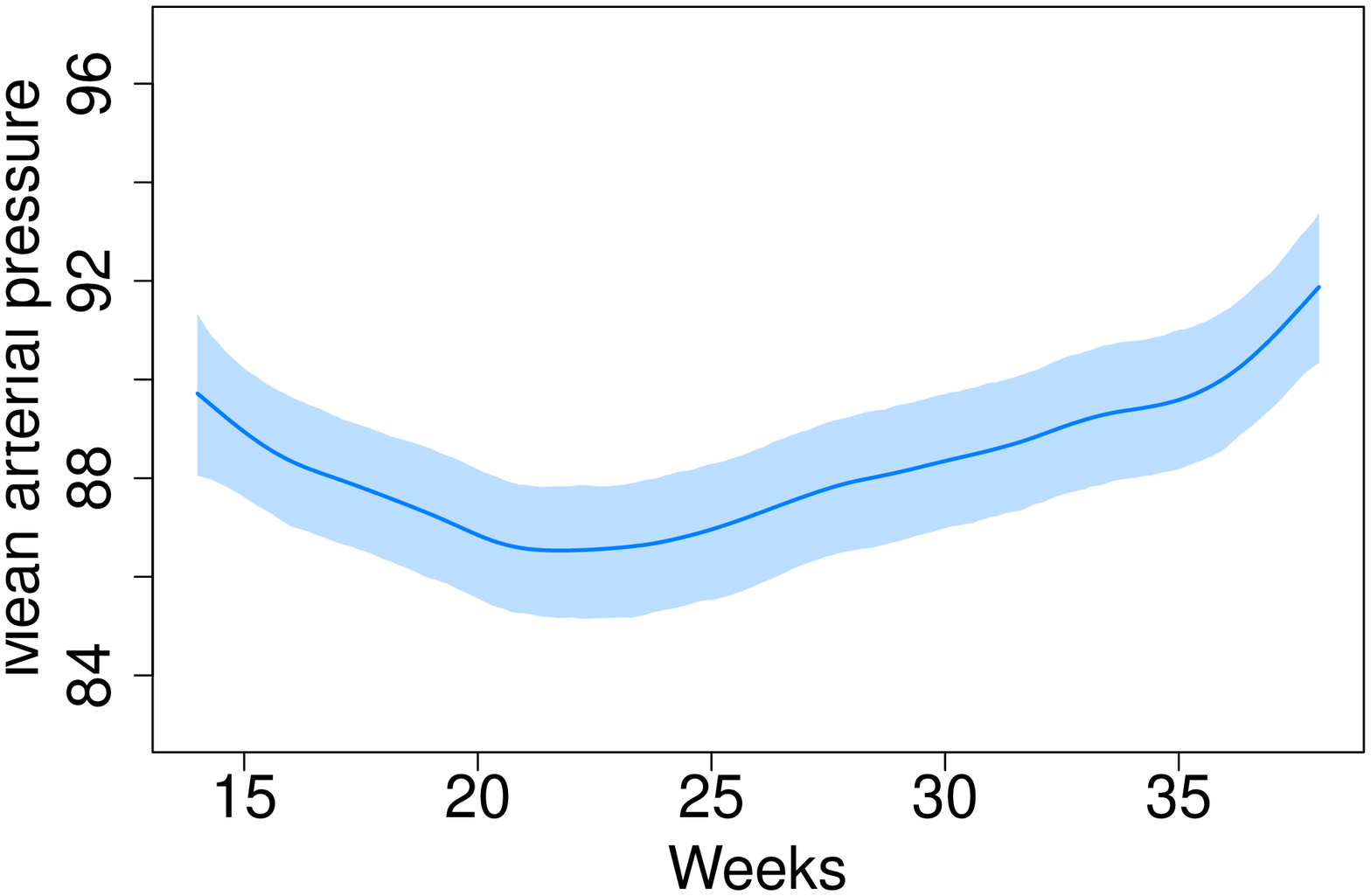}}\\
\subfloat[]{\label{fig:MAPVolPost}\includegraphics[width=0.30\textwidth,angle=270]{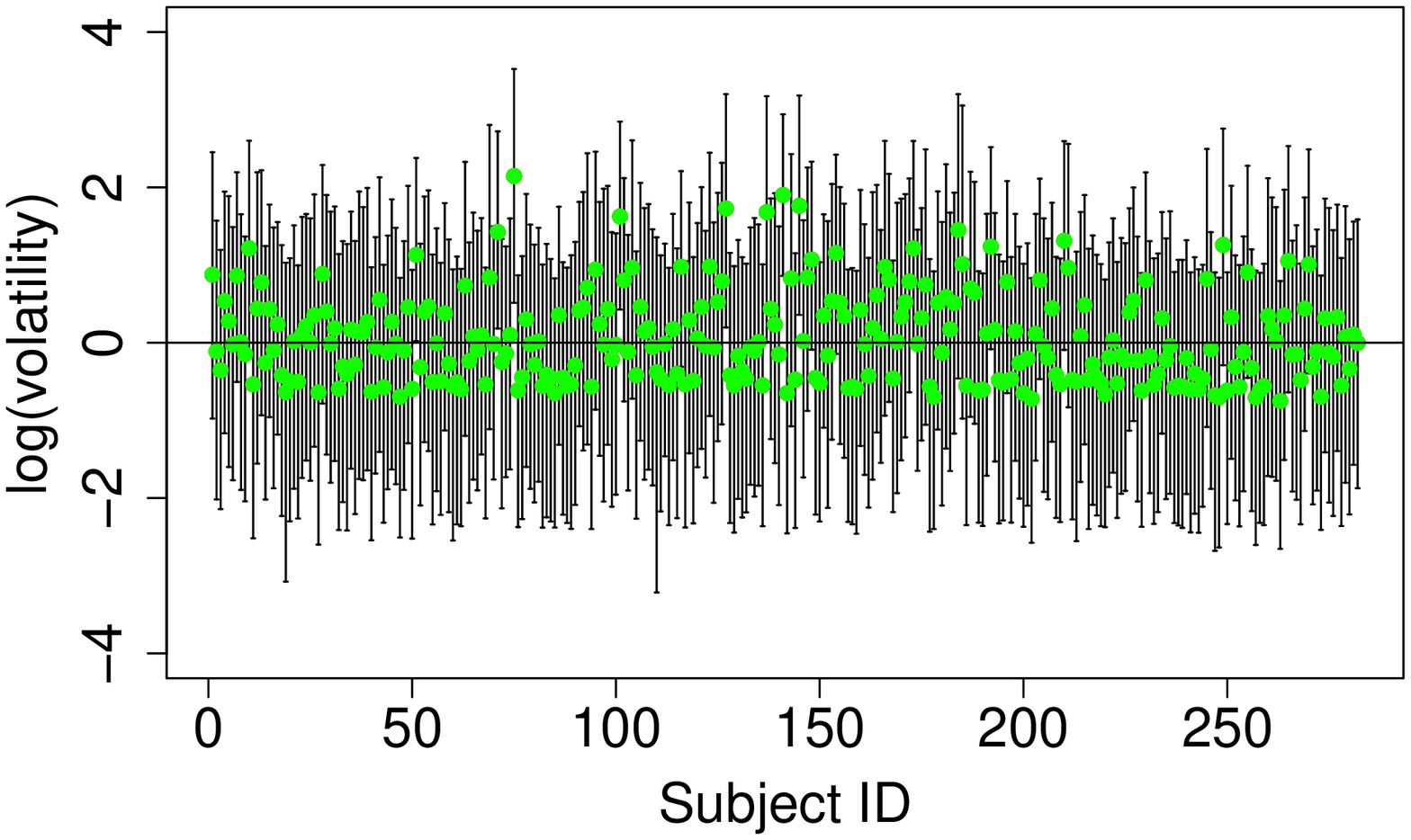}}
\qquad
\subfloat[]{\label{fig:MAPCoeff}\includegraphics[width=0.30\textwidth,angle=270]{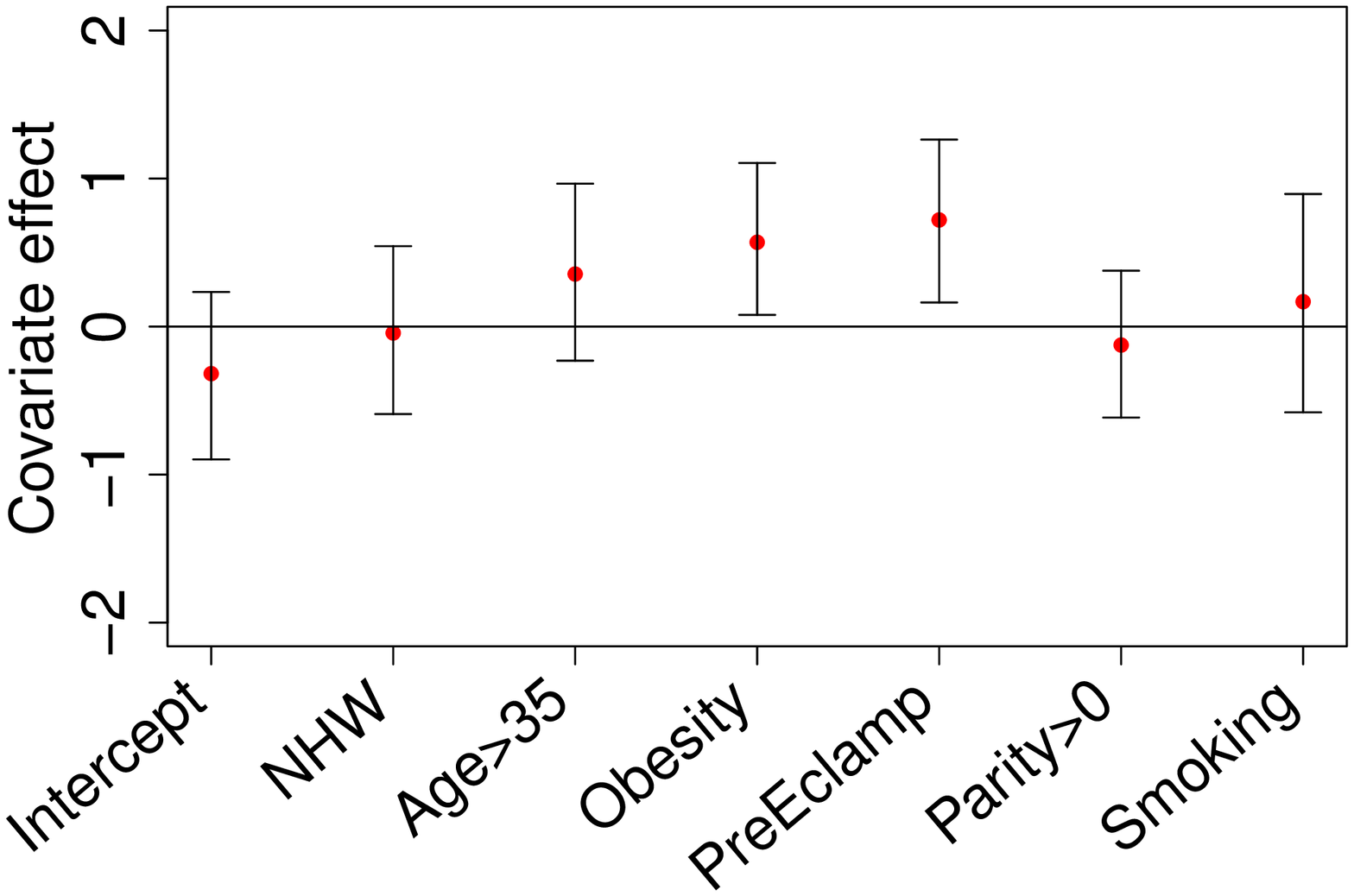}}
\caption{\label{fig:Post} The posterior means and $95\%$  highest posterior density (HPD) credible intervals for (a) the blood pressure during the 2nd and 3rd trimesters for non-Hispanic white group; (b) the blood pressure during the 2nd and 3rd trimesters for non-Hispanic black group; (c) the volatility in the logarithmic scale; (d) covariate effects.}  
\end{figure}

The panels \subref{fig:MAPMNHW} and \subref{fig:MAPMNHB} of Figure \ref{fig:Post} show posterior means and $95\%$ credible intervals of the average blood pressure for NHW and NHB groups respectively, which share a common pattern:  decreasing till the late stage of the second trimester (about 20 to 25 weeks) and then increasing toward the pre-pregnancy level. Within the ethnic group, significant heterogeneity exists in terms of the stability of the blood pressure trajectory at the individual level. As Figure \ref{fig:Post} \subref{fig:MAPVolPost} indicated, posterior means of volatility vary from -0.5 to 2 in the logarithmic scale, suggesting some subjects' trajectories are parallel to the group blood pressure trajectory with very small volatilities while others may significantly depart from the group blood pressure trajectory. 

Most interesting, we find that obesity and preeclampsia are associated with blood pressure volatility, with their $95\%$ credible intervals not covering zero in Figure \ref{fig:Post} \subref{fig:MAPCoeff}. This implies that pregnant women with obesity and/or preeclampsia are more likely to demonstrate irregular patterns of blood pressure relative to their ethnic group. We further examine the characteristics of these subjects with extreme volatilities (results not shown). Among the top eight subjects presenting with the largest volatilities, most of them are NHB with obesity and preeclampsia, do not smoke and give birth to a baby for the first time; half of them are younger than 35. For the eight subjects with the smallest volatilities, they are surprising homogeneous, i.e. all of them being NHW (except one) without obesity and preeclampsia, younger than 35, not smoking and giving birth to a baby before.

\begin{table}[ht]
\caption{Blood pressure data: Posterior summary of parameters in the SVR model.\label{tbl:post}}
\vspace{5pt}
\begin{center}
\begin{tabular}{crrrl}
  \hline\hline
Parameter & Mean & Mode & SD & 95\% HPD invteral \\ 
  \hline
$\sigma^2_\varepsilon$ & 17.807 & 17.818 & 0.694& [16.389, 19.106] \\
$\sigma^2_{M_1}$ & 0.236 & 0.187 & 0.181 & [0.040, 0.556]\\
$\sigma^2_{M_2}$ &0.204 & 0.162 & 0.148 & [0.042, 0.472]\\
$\sigma^2_{U_0}$ & 46.729 & 46.412& 4.776& [38.263, 56.619]\\
$\sigma^2$ & 0.734 & 0.741 & 0.333& [0.082, 1.295] \\
\hline 
\end{tabular}
\end{center}
\end{table}

\section{Discussion}
\label{sec:dis}

We have proposed a stochastic volatility regression model to investigate the volatility and its association with covariates for multi-subject functional data. As an important dynamic feature, volatility measures the stability of the biological process. The analysis of volatility not only reveals its heterogeneity among subjects but also its dependence on the covariates of interest. Complementing the current FDA methods which mainly focus on the trend of trajectory and its derivatives, the SVR method initiates the exploration of stability of functional data.  As illustrated with the blood pressure data, our view is that substantial new insights can be obtained in a rich variety of biomedical applications by studying volatility.     

%The current SVR model can be possibly extended in different directions, based on research interests and characteristics of motivation data sets. For example, we may substitute single group mean function in equation \eqref{eq:obs} by a weighted sum of several mean functions with covariates as the weights and each mean functions as time-varying coefficient. Limited by the sparse observations of HPHB study, we assume the volatility time-constant. Given denser measurements, we may leverage on current methods of stochastic volatility model to allow volatility both vary across time and subjects. As for the regression part, it is also of interest to relieve the normal assumption and to consider high-dimensional covariates.  

\section*{Acknowledgments}
This work was supported by Award Number R01ES017436 and R01ES17240 from the National Institute of Environmental Health Sciences, by funding from the National Institutes of Health (5P2O-RR020782-O3) and the U.S. Environmental Protection Agency (RD-83329301-0), and by the Intramural Research Program of the National Cancer Institute, National Institutes of Health. The content is solely the responsibility of the authors and does not necessarily represent the official views of the National Institute of Environmental Health Sciences, the National Institutes of Health or the U.S. Environmental Protection Agency.

\appendix
\section{Appendix: Proofs of Theoretical Results}
\label{sec:appendix_a}
\subsection{Proof of Theorem \ref{thm:DPSS}}
By the RKHS theory of the polynomial smoothing spline \citep[Section 1.2]{wahba1990spline}, we have
\begin{align*}
{M}_{k}(t) &= {M}_{k0}(t) + {M}_{k1}(t)=
\sum_{l=0}^{p-1} \mu_{kl}\phi_l(t) + \sum_{j=1}^n\nu_{kj}\mathcal{R}_{M_{1}}(t_j, t) +
\eta_{M_1}(t),
\end{align*}
and
\begin{equation*}
\int_{\mathcal{T}}\left\{D^{p}M_k(t)\right\}^2dt =
\bs{\nu}^\prime_{k} \bs{R}_{M_{1}}\bs{\nu}_{k}+ 
\langle\eta_{M_1}(\cdot),\eta_{M_1}(\cdot)\rangle_{\mathcal{H}_{\mathcal{R}_{M_1}}}, 
\end{equation*}
where $M_0(t) \in \mathcal{H}_{\mathcal{R}_{M_0}}$ and $M_1(t) \in \mathcal{H}_{\mathcal{R}_{M_1}}$, the RKHSs $\mathcal{H}_{\mathcal{R}_{M_0}}=\left\{f(t): D^{p} f(t)=0, t \in \mathcal{T}\right\}$ and 
$\mathcal{H}_{\mathcal{R}_{M_1}}=\left\{f(t): D^if(t)\text{ absolutely continuous for } i=0,1,\cdots,p-1, D^{p}f(t) \in \mathcal{L}_2(\mathcal{T}) \right\}$; $\mathcal{R}_{M_0}(s,t)$ and $\mathcal{R}_{M_1}(s,t)$ are reproducing kernels defined in Lemma \ref{lem:GPs_nGP}; 
$\eta_{M_1}(\cdot) \in \mathcal{H}_{\mathcal{R}_{M_1}}$ is orthogonal to $\mathcal{R}_{M_1}(t_j,\cdot)$ with inner product $\langle\mathcal{R}_{M_1}(t_j,\cdot),\eta_{M_1}(\cdot)\rangle_{\mathcal{H}_{\mathcal{R}_{M_1}}}=
\int_{\mathcal{T}}D^p\mathcal{R}_{M_1}(t_j,u) D^p\eta_{M_1}(u)du=0$  for $j=1,2,\cdots,J$; $\mathcal{L}_2(\mathcal{T})=\left\{f(t): \int_{\mathcal{T}}f^2(t)dt < \infty\right\}$ is the space of squared integrable functions defined on index set $\mathcal{T}$.

Similarly,
\begin{align*}
{U}_{i}(t) &= {U}_{i0}(t) + {U}_{i1}(t)=
\sum_{l=0}^{q-1} \alpha_{il}\phi_l(t) +  \sum_{j=1}^{n_i}\gamma_{ij}\mathcal{R}_{U_{1}}(t_{ij}, t)  +
\eta_{U_{i1}}(t),
\end{align*}
and
\begin{equation*}
\int_{\mathcal{T}}\left\{D^{q}U_i(t)\right\}^2dt =
\bs{\gamma}_i^\prime\bs{R}_{U_{i1}}\bs{\gamma}_i+ 
\langle\eta_{U_{i1}}(\cdot),\eta_{U_{i1}}(\cdot)\rangle_{\mathcal{H}_{\mathcal{R}_{U_1}}}, 
\end{equation*}
where $U_{i0}(t) \in \mathcal{H}_{\mathcal{R}_{U_0}}$ and $U_{i1}(t) \in \mathcal{H}_{\mathcal{R}_{U_1}}$ with $\mathcal{H}_{\mathcal{R}_{U_0}}=\left\{f(t): D^{q} f(t)=0, t \in \mathcal{T}\right\}$ and 
$\mathcal{H}_{\mathcal{R}_{U_1}}=\left\{f(t): D^if(t)\text{ absolutely continuous for } i=0,1,\cdots,q-1, D^{q}f(t) \in \mathcal{L}_2(\mathcal{T}) \right\}$ the RKHSs with reproducing kernel $\mathcal{R}_{U_0}(s,t)$ and $\mathcal{R}_{U_1}(s,t)$ defined in Lemma \ref{lem:GPs_nGP}; 
$\eta_{U_{i1}}(\cdot) \in \mathcal{H}_{\mathcal{R}_{U_1}}$ is orthogonal to $\mathcal{R}_{U_1}(t_{ij},\cdot)$ for $j=1,2,\cdots,n_i$.

Hence, the double-penalized sum-of-squares \eqref{eq:DPSS} can be written as\begin{align*}
\textsf{DPSS}=
&\sum_{i=1}^m\frac{1}{n_i} (\bs{Y}_i- \bs{\Delta}_i \bs{\phi}_\mu\bs{\mu}_{k_i} - \bs{\Delta}_i\bs{R}_{M_{1}}\bs{\nu}_{k_i} - \bs{\phi}_{\alpha_i}\bs{\alpha}_i-\bs{R}_{U_{i1}}\bs{\gamma}_i )^\prime \times  \\ 
&(\bs{Y}_i- \bs{\Delta}_i \bs{\phi}_\mu\bs{\mu}_{k_i} - \bs{\Delta}_i\bs{R}_{M_{1}}\bs{\nu}_{k_i} - \bs{\phi}_{\alpha_i}\bs{\alpha}_i-\bs{R}_{U_{i1}}\bs{\gamma}_i )
 + \notag
\\ 
&\sum_{k=1}^g \lambda_{M_k}\bs{\nu}^\prime_{k} \bs{R}_{M_{1}}\bs{\nu}_{k}
+\sum_{i=1}^m\lambda_{U_i}\bs{\gamma}^\prime_i\bs{R}_{U_{i1}}\bs{\gamma}_i+\\
&\sum_{k=1}^g\lambda_{M_k}
\langle\eta_{M_1}(\cdot),\eta_{M_1}(\cdot)\rangle_{\mathcal{H}_{\mathcal{R}_{M_1}}}
+\sum_{i=1}^m\lambda_{U_i}
\langle\eta_{U_{i1}}(\cdot),\eta_{U_{i1}}(\cdot)\rangle_{\mathcal{H}_{\mathcal{R}_{U_1}}},
\end{align*}
which is minimized at $\langle\eta_{M_1}(\cdot),\eta_{M_1}(\cdot)\rangle_{\mathcal{H}_{\mathcal{R}_{M_1}}} =
\langle\eta_{U_{i1}}(\cdot),\eta_{U_{i1}}(\cdot)\rangle_{\mathcal{H}_{\mathcal{R}_{U_1}}} =0$, leading to $\eta_{M_1}(\cdot) =\eta_{U_{i1}}=0$.

\subsection{Proof of Corollary \ref{cor:dpss2}}
Taking partial derivatives of double penalized sum-of-squares in Corollary \ref{thm:DPSS} with respective to $\bs{\mu}_k$, $\bs{\nu}_k$, $\bs{\alpha}_i$ and $\bs{\gamma}_i$ respectively and setting them to zeros, we have
\begin{align*}
\frac{\partial\;\textsf{DPSS}}{\partial\;\bs{\mu}_k}
&=
\displaystyle\sum_{i:k_i=k} \frac{1}{n_i} \bs{\phi}_\mu^\prime\bs{\Delta}_i^\prime
(\bs{\Delta}_i \bs{\phi}_\mu\bs{\mu}_{k_i} + \bs{\Delta}_i\bs{R}_{M_{1}}\bs{\nu}_{k_i} + \bs{\phi}_{\alpha_i}\bs{\alpha}_i+\bs{R}_{U_{i1}}\bs{\gamma}_i-\bs{Y}_i )=\bs{0},\\
\frac{\partial\;\textsf{DPSS}}{\partial\;\bs{\nu}_k}
&=
\displaystyle\sum_{i:k_i=k} \frac{1}{n_i} \bs{R}_{M_1}\bs{\Delta}_i^\prime
(\bs{\Delta}_i \bs{\phi}_\mu\bs{\mu}_{k_i} + \bs{\Delta}_i\bs{R}_{M_{1}}\bs{\nu}_{k_i} + \bs{\phi}_{\alpha_i}\bs{\alpha}_i+\bs{R}_{U_{i1}}\bs{\gamma}_i-\bs{Y}_i )+
\lambda_{M_k}\bs{R}_{M_1}\bs{\nu}_k
=\bs{0},\\
\frac{\partial\;\textsf{DPSS}}{\partial\;\bs{\alpha}_i}
&=
\frac{1}{n_i} \bs{\phi}_{\alpha_i}^\prime
(\bs{\Delta}_i \bs{\phi}_\mu\bs{\mu}_{k_i} + \bs{\Delta}_i\bs{R}_{M_{1}}\bs{\nu}_{k_i} + \bs{\phi}_{\alpha_i}\bs{\alpha}_i+\bs{R}_{U_{i1}}\bs{\gamma}_i-\bs{Y}_i )=\bs{0},\\
\frac{\partial\;\textsf{DPSS}}{\partial\;\bs{\gamma}_i}
&=
\frac{1}{n_i} \bs{R}_{U_{i1}}
(\bs{\Delta}_i \bs{\phi}_\mu\bs{\mu}_{k_i} + \bs{\Delta}_i\bs{R}_{M_{1}}\bs{\nu}_{k_i} + \bs{\phi}_{\alpha_i}\bs{\alpha}_i+\bs{R}_{U_{i1}}\bs{\gamma}_i-\bs{Y}_i ) + \lambda_{U_i}\bs{R}_{U_{i1}}\bs{\gamma}_i=\bs{0},
\end{align*}
which lead to
\begin{align}
\label{eq:d_mu_k}
\bs{\phi}_\mu^\prime\bs{\Delta}\bs{\phi}_\mu\bs{\mu}_k+
\bs{\phi}_\mu^\prime\bs{\Delta}\bs{R}_{M_{1}}\bs{\nu}_{k}
&=\bs{\phi}_\mu^\prime\bs{\tilde{Y}}_k,\\
\label{eq:d_nu_k}
\bs{R}_{M_1}\bs{\Delta}\bs{\phi}_\mu\bs{\mu}_k +
(\bs{R}_{M_1}\bs{\Delta}+\lambda_{M_k}\bs{I})\bs{R}_{M_1}\bs{\nu}_k
&=\bs{R}_{M_1}\bs{\tilde{Y}}_k,\\
\label{eq:d_alpha_i}
\bs{\phi}_{\alpha_i}^\prime\bs{\phi}_{\alpha_i}\bs{\alpha}_i+
\bs{\phi}_{\alpha_i}^\prime\bs{R}_{U_{i1}}\bs{\gamma}_i
&=\bs{\phi}_{\alpha_i}^\prime\bs{\tilde{Y}}_i,\\
\label{eq:d_gamma_i}
\bs{R}_{U_{i1}}\bs{\phi}_{\alpha_i}\bs{\alpha}_i+
(\bs{R}_{U_{i1}}+n_i\lambda_{U_i}\bs{I})\bs{R}_{U_{i1}}\bs{\gamma}_i
&=\bs{R}_{U_{i1}}\bs{\tilde{Y}}_i,
\end{align}
with $\bs{\tilde{Y}}_k=\displaystyle\sum_{i:k_i=k}\frac{1}{n_i}\bs{\Delta}_i^\prime\left(\bs{Y}_i-  \bs{\phi}_{\alpha_i}\bs{{\alpha}}_i-\bs{R}_{U_{i1}}\bs{{\gamma}}_i \right )$, $\bs{\tilde{Y}}_i = \bs{Y}_i- \bs{\Delta}_i \bs{\phi}_\mu\bs{{\mu}}_{k_i} - \bs{\Delta}_i\bs{R}_{M_{1}}\bs{{\nu}}_{k_i}$ and  $\bs{\Delta}= \displaystyle\sum_{i:k_i=k}\frac{1}{n_i}\bs{\Delta}_i^\prime\bs{\Delta}_i $. The solutions of $\bs{\alpha}_i$ and $\bs{\gamma}_i$ in the step (a) can be obtained from equations \eqref{eq:d_alpha_i} and \eqref{eq:d_gamma_i}, while the solutions of $\bs{\mu}_k$ and $\bs{\nu}_k$ in the step (b) from equations \eqref{eq:d_mu_k} and \eqref{eq:d_nu_k}.  

\subsection{Proof of Proposition \ref{prop:IW}}
Based on the SDE $D^p X(t)=\dot{W}(t)$, we have 
\begin{equation*}
D^1 \bs{X}(t)=\bs{C}\bs{X}(t)+\bs{D}\dot{W}(t),
\end{equation*}
where $\bs{X}=\{X(t), D^1X(t),\cdots,D^{p-1}X(t)\}^\prime$, $\bs{C}=(c_{ii^\prime})_{p \times p}$,  $c_{ii^\prime}=1$ when $i^\prime=i+1$ and $c_{ii^\prime}=0$ otherwise, and $\bs{D}=(0, 0, \cdots, 1)^\prime$.

It follows that 
\begin{align*}
\bs{X}_{j+1}&=\exp(\bs{C}\delta_j)\bs{X}_j+
\int_0^{\delta_j}\exp\{\bs{C}(\delta_j-u)\}\bs{D}\bs{\dot{W}}(t_j+u)du\\
&=\bs{G}_j\bs{X}_j+\bs{\omega}_j,
\end{align*}
where $\bs{G}_j=\exp(\bs{C}\delta_j)=\sum_{k=0}^p\frac{\delta_j^k}{k!}\bs{C}^k$ and $\bs{\omega}_{j}  \sim \textsf{N}_p\left(\bs{0}, \bs{W}_j\right)$ with
\begin{align*}
\bs{W}_j&=\int_0^{\delta_j}\exp\{\bs{C}(\delta_j-u)\}\bs{D}\bs{D}^\prime\exp\{\bs{C}^\prime(\delta_j-u)\}du
\end{align*}
as required.

\bibliographystyle{asa}
\bibliography{SVR}

\end{document}